\documentclass[11pt,a4paper]{article}
\usepackage{epsfig}
\usepackage{epic}
\usepackage{float}
\usepackage{afterpage}
\usepackage{amsmath}
\renewcommand{\baselinestretch}{1.2}\normalsize

\newcommand{\be}{\begin{equation}}
\newcommand{\ee}{\end{equation}}
\newcommand{\bea}{\begin{eqnarray}}
\newcommand{\eea}{\end{eqnarray}}
\newcommand{\klgl}{\:\hbox to -0.2pt{\lower2.5pt\hbox{$\sim$}\hss}
{\raise3pt\hbox{$<$}}\:}
\newcommand{\grgl}{\:\hbox to -0.2pt{\lower2.5pt\hbox{$\sim$}\hss}
{\raise3pt\hbox{$>$}}\:}
\begin{document}
%
%
\markboth{ }{ }
\renewcommand{\baselinestretch}{1.1}\normalsize
\vspace*{-2cm}
\hfill TUM-HEP-325/98

\hfill SFB-375-306

\vspace*{2cm}
\bigskip
\bigskip
\begin{center}
{\huge\bf{Thermal Renormalization \\ Group-Equations and the \\Phase-Transition
of \\ Scalar $O(N)$-Theories\\}} 
\end{center}
\bigskip
\begin{center}
{\Large{Bastian Bergerhoff}}\footnote{
bberger@physik.tu-muenchen.de}
{\Large{and J{\"u}rgen Reingruber}}\footnote{
reingrub@physik.tu-muenchen.de}\vspace*{0.3cm}\\
Institut f\"ur Theoretische Physik \\
Technische Universit\"at M\"unchen \\
James-Franck-Strasse, D-85748 Garching, Germany
\end{center}
\setcounter{footnote}{0}
\bigskip
\vspace*{1cm}\begin{abstract}
\noindent
We discuss the formulation of "thermal renormalization group-equations" and
their application to the finite temperature phase-transition of scalar $O(N)$-theories.
Thermal renormalization group-equations allow for a computation of both
the universal and the non-universal aspects of the critical behavior directly
in terms of the zero-temperature physical couplings.
They provide a nonperturbative method for a computation of quantities like 
real-time correlation functions in a thermal environment, where in many
situations straightforward perturbation theory fails due to the bad
infrared-behavior of the thermal fluctuations.
We present results for the critical temperature, critical exponents and
amplitudes as well as the
scaling equation of state for self-interacting scalar theories.
\end{abstract}
\bigskip
\noindent PACS-No: 11.10.Wx, 11.15.Tk, 05.70.Fh, 05.70.Ce
\newpage
\renewcommand{\baselinestretch}{1.2}\normalsize
\section{Introduction}

In recent years, the study of field theories in a thermal environment
has become increasingly popular
\cite{TFT}.
The interest in this field has increased for a number of reasons.
From the experimental side, the study of strong interactions at high
temperatures and densities has been taken up by means of heavy-ion
collisions.
In order to understand theoretically the outcome of such experiments,
we have to learn how to treat field theory in a hot and dense
environment.
Thermal field theory also plays an important r{\^o}le in
astro-particle physics.
Questions concerning the physics of the very early universe can usually
only be addressed if the impact of the temperatures on the
properties of matter is taken into account.
A famous and in recent years extensively discussed question concerns
the phase-transition
that might have been associated with the
restoration of the electroweak symmetry\footnote{We know by now that in the framework of
  the standard model there was no phase-transition
\cite{EWPTreview}.
This does however not exclude phase-transitions in extended models
like the minimal supersymmetric standard model.} at temperatures of the order
of $100$~GeV.
Another interesting phase-transition is the transition from the
quark-gluon plasma phase of quantum-chromodynamics to the low energy
phase where confinement applies and the (approximate) chiral symmetry
of the standard model is broken.
This transition has taken place at temperatures of the order of
$100$~MeV and is expected to be experimentally accessible with the next
generation of heavy-ion colliders.
It is also an interesting open question of particle physics applied to
astronomy whether a quark-gluon plasma is realized in the interior of
neutron stars.

A large number of open problems is connected to non-equilibrium
phenomena.
In the physics of the early universe this concerns for example the
question of reheating after a (hypothetical) inflationary phase.
Also heavy-ion collisions may not yield a thermalized state and one
should strictly speaking consider QCD out of thermal equilibrium.
Finally it is of interest to study the dynamics of phase-transitions
in various contexts.

In marked contrast to the number of situations in which thermal (or
more general non-equilibrium) field theory is of relevance is the
number of open fundamental questions.
This is due to the fact that in many situations even in the simpler
equilibrium case the well established methods of perturbation theory
fail due to a modified infrared-behavior of the theory.
Famous examples of the infrared problems are the observation of
Linde
\cite{Linde},
that the free energy of a non-abelian gauge-theory is not computable in
perturbation theory beyond three-loop order (irrespective of the size
of the coupling constant) or the fact that even super-daisy
resummed one-loop perturbation theory fails to predict the correct critical
behavior of self-interacting scalar theories even if they are weakly coupled
at vanishing temperature
\cite{PertFirstOrder}.
Although the physical mechanisms behind the above-mentioned
problems are quite different, both problems may be traced back to the fact that
the behavior of bulk-observables like the free energy or of theories close to
a second order phase-transition is governed by three-dimensional physics (if
the underlying zero-temperature theory is formulated in 3+1
dimensions).

We then have to devise methods that are able to cope with this
effective change of the relevant degrees of freedom.
One such method in principle is lattice field theory, which may be
used to study time-independent quantities like the free energy also at
non-vanishing temperature.
Lattice studies of field theory at high temperature are however often
complicated by the existence of widely separated scales and there is
as yet no formulation of lattice field theory that can deal with finite
density or time-dependent quantities.

Another way of dealing with theories where the relevant degrees of
freedom are scale-dependent is provided by the renormalization group (RG).
This is a well known fact in statistical physics, where the Wilsonian
form of the renormalization group 
\cite{Wilson}
is widely used in the study of
critical phenomena.
In the context of field theory, there are different implementations of
the general RG idea.
These range from the ``environmentally friendly'' RG
(\cite{ChrisReview} and references therein)
which is a variant of the Callen-Szymanzik RG, via the ``auxiliary mass
method''
\cite{Drummondetal}
to continuum implementations of the Wilsonian approach, initiated in a
field-theory context by Polchinski and others
\cite{Polchinski}.
This approach has in recent years been applied to a number of problems
in field theory both at vanishing and non-vanishing temperature and is
now also known under the name ``Exact renormalization
group (ERG)''-approach.
The ERG-approach is a nonperturbative formulation of field
theory and by construction avoids any infrared-problems irrespective
of the dimensionality of the system under study.
It is formulated in Euclidean space and may be straightforwardly
adapted to study thermal field theory in the ``imaginary-time''- or
Matsubara formalism.

Recently, an implementation of the Wilsonian renormalization group in
the real-time formulation of thermal field theory has been proposed
\cite{DAP1}.
This ``thermal renormalization group'' (TRG) has a number of
advantages as compared to the ERG applied to thermal field theory.
Even though the formulations of field theory in the imaginary- and the
real-time formalism are in principle equivalent and one can compute
any quantity of interest in both approaches, in practice recovering
Green-functions at real time arguments from their expressions at
discrete imaginary energies through analytical continuation is often
tedious.
Moreover, in practical applications of the Wilsonian RG results will
usually be obtained numerically, making analytical continuation
impossible.
Thus, in order to study quantities like damping rates and the like
\cite{Massimo}, 
a real-time formulation of thermal field theory is mandatory.

Also, in the real-time formalism there is a clear separation between
``thermal'' and ``quantum'' fluctuations.
The thermal renormalization group only treats the thermal fluctuations
of the theory, all quantum fluctuations are assumed to be integrated
out, i.e.~one starts with the full physical effective action of the
theory at vanishing temperature.
This feature essentially prevents us from studying theories that are
strongly interacting already at $T=0$ and may thus seem like a major
shortcoming.
On the other hand, in situations where the zero-temperature theory is
weakly coupled this apparent shortcoming turns into an advantage.
In such a situation the calculation of non-universal quantities such
as critical temperatures in terms of the measurable
couplings of the theory in the framework of the ERG requires a two
step procedure, where one first has to relate physical quantities at
zero temperature to the parameters of the action. 
These parameters are renormalization scheme dependent and one has to
perform perturbative calculations using the ERG-formulation
\cite{UliSchemes}.
In situations where two-loop accuracy is required to fix the
ambiguities of such a calculation, this procedure is highly
nontrivial and has up to now only been performed in simple models
\cite{ThomasChristof}.

Another advantage of a formulation of RG-equations only for the
thermal modes is the fact that this can be done while respecting
manifest gauge-invariance
\cite{DAP2}.
The general formulation of Wilsonian RG-equations relies on a
separation of ``hard'' and ``soft'' modes by means of an external
scale which acts as a momentum cutoff.
This procedure is in general not consistent with
gauge-invariance\footnote{There is a formulation using the
  background-field approach where one can keep manifest
  gauge-invariance with respect to background-gauge transformations
\cite{MC1}.} (see however
\cite{SBLiao} for a different approach circumventing this problem),
and one obtains generalized Slavnov-Taylor identities that may be
interpreted as fine-tuning conditions that have to be respected in
order to regain BRS-invariance of the physical effective action
\cite{MSTI}.
In the framework of the TRG one avoids this problem since only the
physical fields have thermal fluctuations.
Manifest gauge-invariance severely restricts the form of possible
contributions to the effective action and thus helps in making
sensible approximations (see the discussion below).

In the present paper we present an extensive application of the TRG to
the critical behavior of the simplest nontrivial
field-theories in $3+1$ dimensions, scalar self-interacting models
with a (at zero temperature spontaneously broken) $O(N)$-symmetry.
Despite their simplicity, already these theories are
not accessible in straightforward perturbation theory for
temperatures close to the critical one.
$O(N)$-symmetric scalar theories have a number of applications both in
the context of statistical physics and in particle physics.
In statistical physics, these models are used for example for the description
of polymers (in the limit $N=0$), the liquid-vapor transition ($N=1$, the
Ising model), the transition of helium to a superfluid state ($N=2$) or
ferromagnetic systems (the Heisenberg model, $N=3$).
In particle physics, the model with $N=4$ at zero temperature
describes the scalar part of the Lagrangian of the standard model.
At non-vanishing temperature the theory with one scalar field shows
the same universal behavior as the standard model for the critical
value of the zero-temperature Higgs mass
\cite{EWPTsecondorder}.
For $N=4$ the universal behavior should be the same as that of the
chiral phase-transition in the $N_f=2$ chiral limit of QCD.

The universal behavior of the theory may be studied directly in three
dimensions.
There are accurate results on critical exponents from methods like
(Borel-resummed) $\epsilon$-expansion, from perturbative calculation
at fixed dimension, from resummed high-temperature series or from
lattice studies
(for a review, see 
\cite{ZJ}).
There are also a number of studies using the ERG directly in three
dimensions
(for example \cite{ChristofNickCritExp,JNC1,CS}).
Also the scaling equation of state (EOS) has been widely studied in the
three-dimensional theory
(\cite{JNC1}-\cite{Tsypin}, see also \cite{ZJ}).

On the other hand, the number of studies of the full $3+1$-dimensional
theory at finite temperature close to the phase-transition is rather
limited.
Results on the critical temperature and exponents in the ERG-approach
are given in 
\cite{ChristofNickfiniteT}-\cite{Mike2},
and for $N=1$ 
$T_c$, exponents and critical amplitude ratios have been studied by
means of the environmentally friendly RG in 
\cite{Chrisetal1}.
The critical temperature and some exponents have also been
obtained through the auxiliary mass method in
\cite{Satoetal}.
Results for critical exponents from the TRG have been given for $N=1$
in 
\cite{DAP1}
and critical exponents as well as the equation of state for $N=1$ have
been obtained from the TRG in 
\cite{B1}.
We here extend these studies to arbitrary $N$ and furthermore give
critical temperatures, the scaling form of the equation of state as
well as critical amplitudes in terms of the zero-temperature couplings for the
cases $N=1$ and $N=4$.
For reasons to be discussed below our results on the universal
critical behavior are not as accurate as the existing results
obtained in the effective three-dimensional theory.
On the other hand, the present work does not rely on universality
arguments and the approximations made in the present paper may be improved.
The TRG allows for the direct computation also of real-time quantities and
offers a convenient nonperturbative tool to study field theory in a hot and --
after a straightforward extension of the method possibly dense -- environment.

The format of the paper is as follows:
In the next section we give a review of the formulation of the thermal
renormalization group-equation and discuss some general points regarding the
possibilities of approximately solving equations of this general type.
Here we can rely on experience gained from the study of the
Euclidean exact renormalization group-equation.
In section 3 we apply the method to $O(N)$-symmetric scalar
theories.
After deriving the flow-equation for the effective potential we discuss in
detail the phase-transition for $N=1$ and $N=4$.
We compare our results for the critical temperatures, critical
exponents and amplitudes and the scaling equation of state with
different results given in the literature.
We also discuss the dependence of the critical temperature on the number of
fields and compare with the expectations from large-$N$ and naive perturbation theory.

Section 4 contains our concluding remarks.

\section{The Thermal Renormalization Group-equation}

In this section we discuss the derivation of the thermal renormalization group
(TRG)-equation 
\cite{DAP1}
and some more general points on its relevance and strategies for
its approximate solution.

The basic objective of thermal renormalization group-equations in the framework
proposed in 
\cite{DAP1} is to use the Wilsonian RG-approach in the real-time formulation of
thermal field theory in order to have access to real-time correlation functions
avoiding the need for analytical continuation.
Wilsonian renormalization group-equations are generally not very useful in
Minkowski-space, since the connection between $k^2$ and "softness" of modes is lost.
On the other hand, in many situations the problematic behavior of perturbation
theory at finite temperature is only due to the thermal modes, and the dynamics
of the theory in the vicinity of a phase-transition is governed by
three-dimensional classical statistics. 
To deal with this particular problem, one may use the fact that in real-time
the propagators clearly discriminate between "thermal" and "quantum"
fluctuations to treat the thermal contributions with renormalization group
methods, leaving the quantum contributions untouched. 

\subsection{Formulation of the TRG-equation}

Field-theory, not necessarily in a thermal equilibrium situation, may
be formulated using the Schwinger-Keldysh or closed time-path (CTP) formulation
\cite{CTP}.
Here one is interested in expectation-values of operators with
respect to some general density-matrix rather than in scattering amplitudes,
i.e. matrix-elements of 
the form $\langle{\mbox{out}}|{\mbox{in}}\rangle$ as in the more conventional
approach to field-theory at vanishing 
temperature.
In the CTP-formalism one thus considers matrix
elements of the form $\langle{\mbox{in}}|{\mbox{in}}\rangle$.
In order to be able to derive correlation-functions at different times, one defines the generating functional $Z$ which should now depend on two
sources (in the case of one real field) $J_1$ and $J_2$ by inserting a complete
set of states at some time $t$ as
\bea
Z[J_1,J_2] &=& \int\delta \Phi \langle{\mbox{in}},t_0|\Phi,t\rangle_{J_2}
\langle\Phi,t|{\mbox{in}},t_0\rangle_{J_1}
\label{ZJ1J2}
\eea
where all states are in the Heisen\-berg picture and $Z$ of course depends on the
initial state of the system. 
One may as usually introduce the path-integral representation for the
transition matrix elements $\langle{\mbox{in}},t_0|\Phi,t\rangle_{J_i}$ to find
\bea
Z[J_1,J_2] &=& \int \delta \Phi \langle{\mbox{in}},t_0|T^* \exp\left(-i \int_{t_0}^t d\bar{t}
d^3x J_2 \phi_H \right) | \Phi,t\rangle \times \nonumber \\
&& \qquad \quad \times \langle\Phi,t|T \exp\left(i \int_{t_0}^t d\bar{t} d^3x J_1 \phi_H \right)
|{\mbox{in}},t_0\rangle
\label{ZJ1J2pi}
\eea
where $T$ and $T^*$ represent the time- and the anti-time-ordering operator
respectively and $\phi_H$ are Heisen\-berg-fields.
Note that the sources $J_1$ and $J_2$ have to be different in order to obtain
time-dependent correlation functions.

Finally, one may also formulate the generating functional by introducing the
density matrix $\rho$ that characterizes the initial state. 
The generating functional then simply is the
ensemble-average of the product of the time- and anti-time-ordered
exponentials and we can write
\bea
Z[J_1,J_2,\rho] &=& N {\mbox{ Tr}}\biggl\lbrace \rho \left[ T^* \exp\left(-i \int_{t_0}^t d\bar{t}
d^3x J_2 \phi_H \right) \right] \times \nonumber \\
&& \qquad \times \left[ T \exp\left(i \int_{t_0}^t d\bar{t} d^3x J_1 \phi_H \right)
\right] \biggr \rbrace \nonumber \\
&=& \int \delta \phi \delta \phi' \delta \Phi \langle\phi,t_0|\rho |\phi',t_0\rangle \langle\phi',t_0| T^*
\exp\left(-i \int_{t_0}^t d\bar{t} 
d^3x J_2 \phi_H \right) |\Phi,t\rangle \times \nonumber \\
&& \qquad \quad \times \langle\Phi,t| T \exp\left(i \int_{t_0}^t d\bar{t} d^3x J_1 \phi_H,t_0
\right) | \phi, t_0 \rangle
\label{Zrho}
\eea
Derivatives of $Z$ with respect to the sources thus generate ensemble-averages
(expectation values) of products of Heisenberg-fields.
Due to the introduction of the two independent sources and the fact that we are
computing matrix-elements between "in"-states at equal time, while inserting a
complete set of states at a different time $t$ and using time- and
anti-time-ordered exponentials the integration-contour is a closed path from $t_0$
to $t$ and back, thus giving rise to the name "closed-time-path" formulation
(figure 1).
\begin{figure}
\begin{center}
\begin{minipage}[b]{.8\linewidth}
\centering\epsfig{file=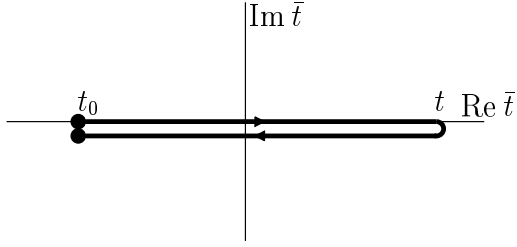}
\renewcommand{\baselinestretch}{1}\normalsize
{\footnotesize{\footnotesize\caption{The Schwinger-Keldysh contour in the complex time-plane.
The lower part of the contour is infinitesimally shifted below the real
axis.}}}
\end{minipage}
\end{center}
\end{figure}
\renewcommand{\baselinestretch}{1.2}\normalsize
Finally we go over to a path-integral representation of the transition
elements and write
\bea
Z[J_1,J_2,\rho] &=& \int \delta \phi_1 \delta \phi_2 \langle\phi_1,t_0|\rho
|\phi_2,t_0\rangle \exp \left[ i \left( S[\phi_1] - S^*[\phi_2] + J_1 \phi_1 - J_2 \phi_2\right) \right]
\label{Zrho2}
\eea

Up to now, the initial state is completely unspecified. 
Indeed, the matrix element $\langle \phi_1 | \rho | \phi_2 \rangle$ will be
some functional of the configurations $\phi_i(\vec{x},\bar{t}=t_0)$ which may be
written as 
\bea
\langle \phi_1,t_0 | \rho | \phi_2,t_0 \rangle &=& \exp \left( i K[\phi_1,\phi_2] \right)
\nonumber \\
&=& \exp \left[ i \left(K + \int d^3k K^a \phi_a + \frac{1}{2}\int d^3k d^3p K^{ab} \phi_a
\phi_b + \ldots \right) \right]
\label{rhoexp}
\eea
where we use a compact notation with $a,b = 1,2$ and metric $c_{ab} =
{\mbox{diag}}(+1,-1)$.
We may then treat the coefficients $K^{ab...}$ as sources for the corresponding
composite operators and write the generating functional as
\bea
Z[J^a,K,K^a,K^{ab},...] \!=\!\! \int \!\!\delta \phi_1 \delta \phi_2 \exp\left[i \left(
S[\phi_1] \!-\! S^*[\phi_2] \!+\! J^a \phi_a \!+\! K \!+\! K^a \phi_a \!+\! \frac{1}{2} K^{ab}
\phi_a \phi_b \!+ ... \right) \right]
\label{ZJK}
\eea
Having specified the density matrix, i.e. the sources $K^{ab...}$, one may then proceed along the lines of 
\cite{CJT}
in order to obtain e.g. a perturbative series etc.

We will in the present paper be interested in systems with a density matrix of
the general form 
\bea
\rho_f = C \exp\left( - \int \frac{d^3k}{(2\pi)^3 2 \omega_k} f(|\vec{k}|) a^\dagger_k a_k \right)
\label{rhogen}
\eea
where $a^\dagger$ and $a$ are creation- and annihilation-operators of a free
scalar field respectively.
(\ref{rhogen}) is a slight generalization of the corresponding thermal
density-matrix which obtains for $f(|\vec{k}|) = \beta \omega_k$ (we may
interpret (\ref{rhogen}) as the density-matrix of a system in which each mode
of the field is at a different temperature given by $T_k =
\frac{\omega_k}{f(|\vec{k}|)}$ --- we will come back to this interpretation
below).
For the above density-matrix the matrix-elements
of $\langle \phi_1 | \rho | \phi_2 \rangle$ are given by
\cite{CalzettaHu37}
\bea
\langle \phi_1 | \rho_f | \phi_2 \rangle &=& C \exp \biggl \lbrace -
\frac{1}{2} \int \frac{d^3k}{(2\pi)^3} \frac{\omega_k}{\sinh f(|\vec{k}|)} 
\left[ \left( \phi_1^2 + \phi_2^2 \right) \cosh
f(|\vec{k}|) - 2 \phi_1 \phi_2 \right] \biggr \rbrace
\label{rhogen12}
\eea
We observe that $K^a$ as well as $K^{abc}$ and all higher derivatives of
$K[\phi]$ vanish in this case. 
The constant is absorbed in the normalization of the generating functional.
The fact that only terms quadratic in $\phi_a$ appear in the density-matrix
element is of course the reason why in finite temperature field-theory in
equilibrium only the two-point functions are modified. 
Indeed, noting that $K^{ab}$ only has support at $t=t_0$, we may absorb the
effects of the initial state into the boundary-conditions for the free
field-equations, where they simply modify the two-point functions, yielding in
the case of a thermal initial state the
well-known real-time propagators of the CTP-formalism at finite temperature.

Above we have made use of the path-integral approach to the CTP-formalism.
In simple theories, a convenient way to obtain explicit expressions for the
propagators is through the operator-formulation.
To this end, introduce a generalization of the $\Theta$-function with real
time-arguments to a $\Theta$-function on the contour $C$ corresponding to
eq.~(\ref{Zrho}) and evaluate the path-ordered product ($T_c$ here is the
path-ordering operator)
\bea
G_f^{(c)}(x,x') &=& \langle T_c \phi(x) \phi(x') \rangle_f \nonumber \\
&=& \Theta_c(t-t')\langle\phi(x) \phi(x') \rangle_f + \Theta_c(t'-t) \langle
\phi(x') \phi(x) \rangle_f
\label{Prop}
\eea
after substituting for the field operators 
\bea
\phi(x) = \int \frac{d^3k}{(2\pi)^3 2 \omega_k} \left( a_k e^{-ikx} +
a^\dagger_k e^{ikx} \right)
\label{freefield}
\eea
In (\ref{Prop}), the averages are ensemble-averages with respect to the
density-matrix defined in (\ref{rhogen}) and in the usual manner one obtains,
using
\bea
\langle a^\dagger_k a_p \rangle_f &=& (2\pi)^3 2 \omega_k
\frac{1}{e^{f(|\vec{k}|)}-1} \delta(\vec{k}-\vec{p}) \nonumber \\
\langle a_k a^\dagger_p \rangle_f &=& (2\pi)^3 2 \omega_k
\left(1+\frac{1}{e^{f(|\vec{k}|)}-1}\right) \delta(\vec{k}-\vec{p})
\label{faverages}
\eea
(note that these expressions again reduce to the well-known results for
$f(|\vec{k}|) = \beta \omega_k$),
\bea
G_f^{(c)}(x-x') &=& i D_f^{(c)}(x-x') = \int \frac{d^4k}{(2\pi)^4} 2 \pi
\delta(k^2-m^2) e^{-ik(x-x')} \times \nonumber \\
&& \qquad \quad \times \left[ \Theta(k_0) \Theta_c(t-t') + \Theta(-k_0)
\Theta_c(t'-t) + \left( \exp(f(|\vec{k}|)-1 \right)^{-1} \right]
\label{Propexp}
\eea
Returning to the notation in terms of two fields $\phi_1$ and $\phi_2$ and
noting that the time-contour starts at $t_0 \rightarrow -\infty$ and goes to $t
\rightarrow + \infty$ and back shifted by an infinitesimal amount below the real
axis (see figure 1), and thus $\Theta_c$ obeys
\bea
\Theta_c(t-t') &=& \Theta(t-t') \qquad {\mbox{on the upper part of $C$}} \nonumber \\
&=& \Theta(t'-t) \qquad {\mbox{on the lower part of $C$}} \nonumber \\
&=& 0 \qquad {\mbox{for $t$ on the upper and $t'$ on the lower part of $C$}}
\nonumber \\
&=& 1 \qquad {\mbox{for $t$ on the lower and $t'$ on the upper part of $C$}}
\label{thetac}
\eea
we can write the propagator as a $2 \times 2$-matrix in the form (note that our
sign-conventions are slightly different from those in 
\cite{DAP1})
\bea
D_f(k) &=& \left(
\begin{array}{cc} \Delta_0 + (\Delta_0 - \Delta^*_0) N_f(|\vec{k}|) &
-(\Delta_0 - \Delta^*_0) \left( \Theta(-k_0) + N_f(|\vec{k}|) \right) \\
-(\Delta_0 - \Delta^*_0) \left( \Theta(k_0) + N_f(|\vec{k}|) \right) &
- \Delta^*_0 + (\Delta_0 - \Delta^*_0) N_f(|\vec{k}|) \end{array} \right)
\label{matprop}
\eea
where we have already switched to momentum-space and use
\bea
\Delta_0(k^2) = \frac{1}{k^2-m^2+i\epsilon}
\label{Delta0}
\eea
and
\bea
N_f(|\vec{k}|) = \frac{1}{e^{f(|\vec{k}|)}-1}
\label{distf}
\eea

We are now in the position to formulate the renormalization group-equations.
As stated above, the general idea of Wilsonian or exact renormalization group
(RG)-equations is to introduce an external dimensionful parameter and divide
the path-integral into a part that includes only the hard modes and a remainder
including the soft modes, where the separation is done with respect to the
external scale.
The result of the path-integral over the hard modes is then treated as an
effective action. 
Instead of performing any of these path-integrals, one derives functional
differential equations for the dependence of the effective action
$\Gamma_\Lambda$ on the
external scale (we shall denote it by $\Lambda$) and solves these equations
for $\Lambda \rightarrow 0$.
In this way one recovers the full solution to the path-integral and thus the
solution of the theory.
Since at any point one integrates only over an infinitesimal "shell" of momenta
between $\Lambda$ and $\Lambda + d \Lambda$, the resulting equations are
formally one-loop.

We have also pointed out that the formulation of exact renormalization
group-equations for a theory in Minkowski-space at vanishing temperature is
possible but not very useful.
The reason is that the Lorentz-invariant separation of modes with $k^2 >(<)
\Lambda^2$ as being hard (soft) has no well-defined meaning.
This translates to the fact that we can give no simple starting-value for the solution
of the RG-equations as $\Lambda \rightarrow \infty$.
In contrast, in Euclidean space and for a conveniently defined
$\Lambda$-dependent effective action the boundary condition is simply that
$\Gamma_\Lambda$ becomes the renormalized classical action of the theory as
$\Lambda \rightarrow \infty$
\cite{Christof}.
To treat field theories in Minkowski-space with Wilsonian RG-equations, we would
have to use cutoffs that break Lorentz-invariance.

The situation is different at finite temperature where Lorentz-invariance is
broken anyway by the presence of the heath-bath with a preferred frame of
reference.
We could thus introduce cutoffs that, e.g., treat $k_0$ and $\vec{k}$
differently.
In an imaginary-time formulation this strategy is pursued in
\cite{Mike2}.
On the other hand, most of the problems of perturbative calculations in thermal
field theory are actually related only to the infrared-behavior of the
thermal modes. 
Since there is a clear separation in the two-point function between thermal and
quantum contributions (cf.~eq.~(\ref{matprop})) and the thermal fluctuations are
on-shell (note that as $\epsilon \rightarrow 0$ we have $\Delta_0 - \Delta_0^*
\rightarrow -2 \pi i \delta(k^2-m^2)$), and thus are effectively depending on
the (Euclidean) space-components of $k$ alone, we may treat these
fluctuations with renormalization-group methods.

We proceed as follows: As noted above, with the general form of the
density matrix given in (\ref{rhogen}) we may control the boundary conditions
for the fields. 
We have seen that thermal boundary conditions correspond to 
\bea
f(|\vec{k}|) = \beta \omega_k = \beta \sqrt{|\vec{k}|^2 + m^2}
\label{fthermal}
\eea
If we want the "hard thermal" modes with $|\vec{k}| \gg \Lambda$ to be in thermal equilibrium,
we should thus introduce the external scale into the function $f$ such that 
\bea
f_\Lambda(|\vec{k}|) \rightarrow \beta \omega_k \quad {\mbox{as $|\vec{k}| \gg
\Lambda$}}
\label{fLambdagg}
\eea
On the other hand, we note that the thermal contributions in the two-point
function (\ref{matprop}) are absent in the limit $f(|\vec{k}|) \rightarrow
\infty$ (in the thermal case this would simply correspond to $T \rightarrow
0$).
Hence we suppress thermal fluctuations with $|\vec{k}| \ll \Lambda$ by requiring
\bea
f_\Lambda(|\vec{k}|) \rightarrow \infty \quad {\mbox{as $|\vec{k}| \ll
\Lambda$}}
\label{fLambdall}
\eea
Having thus introduced the external scale $\Lambda$ we now proceed to derive
the renormalization group-equations. 
The easiest way to do this is to turn back to a path-integral representation of
the generating functional $Z_\Lambda$ in the presence of the modified density-matrix.
Using the inverse of the propagator-matrix $D_f$, which we now denote as
$D_\Lambda^{-1}$ to stress the dependence on the external scale, 
\bea
D_\Lambda^{-1} = \frac{1}{\Delta_0 \Delta^*_0} \left(
\begin{array}{cc} \Delta^*_0 - (\Delta_0 - \Delta^*_0) N_f(|\vec{k}|) &
-(\Delta_0 - \Delta^*_0) \left( \Theta(-k_0) + N_f(|\vec{k}|) \right) \\
-(\Delta_0 - \Delta^*_0) \left( \Theta(k_0) + N_f(|\vec{k}|) \right) &
- \Delta_0 - (\Delta_0 - \Delta^*_0) N_f(|\vec{k}|) \end{array} \right)
\label{DLambdainv}
\eea
we have
\bea
Z_\Lambda[J_1,J_2] &=& \int \delta \phi_1 \delta \phi_2 \exp\left[ i
\left( \frac{1}{2} \phi_a \left( D_\Lambda^{-1}\right)^{ab} \phi_b +
S_{\mathrm{int}}[\phi_1,\phi_2] + J^a \phi_a \right) \right]
\label{ZLambda}
\eea
where $S_{\mathrm{int}}$ is the interaction-part of the classical action with
$S_{\mathrm{int}}[\phi_1,\phi_2] = S_{\mathrm{int}}[\phi_1] - S_{\mathrm{int}}^*[\phi_2]$.
It is straightforward to derive an equation for the dependence of
$Z_\Lambda$ on $\Lambda$.
One finds
\cite{DAP1}
\bea
\Lambda \frac{\partial}{\partial \Lambda} Z_\Lambda[J] = 
- \frac{i}{2} {\mbox{Tr}} \left[ \frac{\delta}{\delta J} \left( \Lambda
  \frac{\partial}{\partial \Lambda} D_\Lambda^{-1} \right) \frac{\delta}{\delta J}
  Z_\Lambda[J] \right]
\label{dLambdaZ}
\eea
where here and below we use the compact notation $J = (J_1,J_2)$ and
correspondingly for the fields.
Defining as usually the generating functional for the connected Green functions
$W_\Lambda[J] = - i \ln Z_\Lambda[J]$ we obtain the RG-equation for
$W_\Lambda$,
\bea
\Lambda \frac{\partial}{\partial \Lambda} W_\Lambda[J] = - \frac{i}{2}
{\mbox{Tr}} \left[ \left( \Lambda \frac{\partial}{\partial \Lambda}
D_\Lambda^{-1} \right)
\frac{\delta^2 W_\Lambda[J]}{\delta J \delta J} \right] + 
\frac{1}{2} {\mbox{Tr}} \left[ \frac{\delta W_\Lambda[J]}{\delta J} \left( \Lambda
\frac{\partial}{\partial \Lambda} D_\Lambda^{-1} \right) \frac{\delta
W_\Lambda[J]}{\delta J} \right]
\label{dLambdaW}
\eea
Finally, we define the generating functional of 1PI Green functions as the
Legendre transform of $W_\Lambda$:
\bea
\Gamma_\Lambda[\varphi] = W_\Lambda[J] - J^a \varphi_a -
\frac{1}{2} \varphi_a \left( D_\Lambda^{-1} \right)^{ab} \varphi_b
\label{GammaL}
\eea
where 
\bea
\varphi_a = \frac{\delta W_\Lambda[J]}{\delta J^a}
\label{vphi}
\eea
and we have subtracted a term bilinear in the field, corresponding to the free
(cutoff-)propagator.
In the RG-equation for $\Gamma_\Lambda$, subtracting the free propagator leads
to a term that cancels the contribution from the second term on the right hand side
of (\ref{dLambdaW}) and we obtain the thermal renormalization group-equation
\cite{DAP1}
\bea
\Lambda \frac{\partial}{\partial \Lambda} \Gamma_\Lambda[\varphi] = \frac{i}{2}
{\mbox{Tr}} \left[ \Lambda \frac{\partial}{\partial \Lambda} D_\Lambda^{-1}
\left( D_\Lambda^{-1} + \frac{\delta^2 \Gamma_\Lambda[\varphi]}{\delta \varphi
\delta \varphi} \right)^{-1} \right]
\label{TRG}
\eea
Finally we have to discuss the limits $\Lambda \rightarrow \infty$ and $\Lambda
\rightarrow 0$.
Let us start with the limit $\Lambda \rightarrow 0$.
In this case, the density-matrix according to (\ref{rhogen}) and
(\ref{fLambdagg}) is the usual density-matrix for a system at thermal
equilibrium with temperature $\beta^{-1}$. Accordingly, the inverse propagator 
$D_\Lambda^{-1}$ becomes the free thermal propagator
(cf.~eq.~(\ref{DLambdainv})) and the effective action tends to 
\bea
\lim_{\Lambda \rightarrow 0} \Gamma_\Lambda[\varphi] = \Gamma[\varphi] -
\frac{1}{2} \varphi_a\left(D^{-1}\right)^{ab} \varphi_b
\label{Lambdatozero}
\eea
where (\ref{GammaL}) was used.

On the other hand, to consider the limit $\Lambda \rightarrow \infty$, note
that $\lim_{\Lambda \rightarrow \infty} N_f = 0$ according to (\ref{distf}) and
(\ref{fLambdall}).
In this limit, the propagator is the one of the theory at vanishing
temperature (cf.~eq.~(\ref{matprop})).
The starting value of the flow-equation as $\Lambda \rightarrow \infty$ is thus
the effective action of the CTP-formalism at vanishing temperature.

Before proceeding to an application of the formalism, let us add some general remarks.
First, note that even though equation (\ref{TRG}) is an exact result, it may
be formally considered as a one loop-equation. 
The corresponding diagram is given in figure 2, where the squared cross denotes
the insertion of $\Lambda \partial_\Lambda D_\Lambda^{-1}$ and the line with
the dark dot is the full propagator.
\begin{figure}
\begin{center}
\begin{minipage}[b]{.8\linewidth}
\centering\epsfig{file=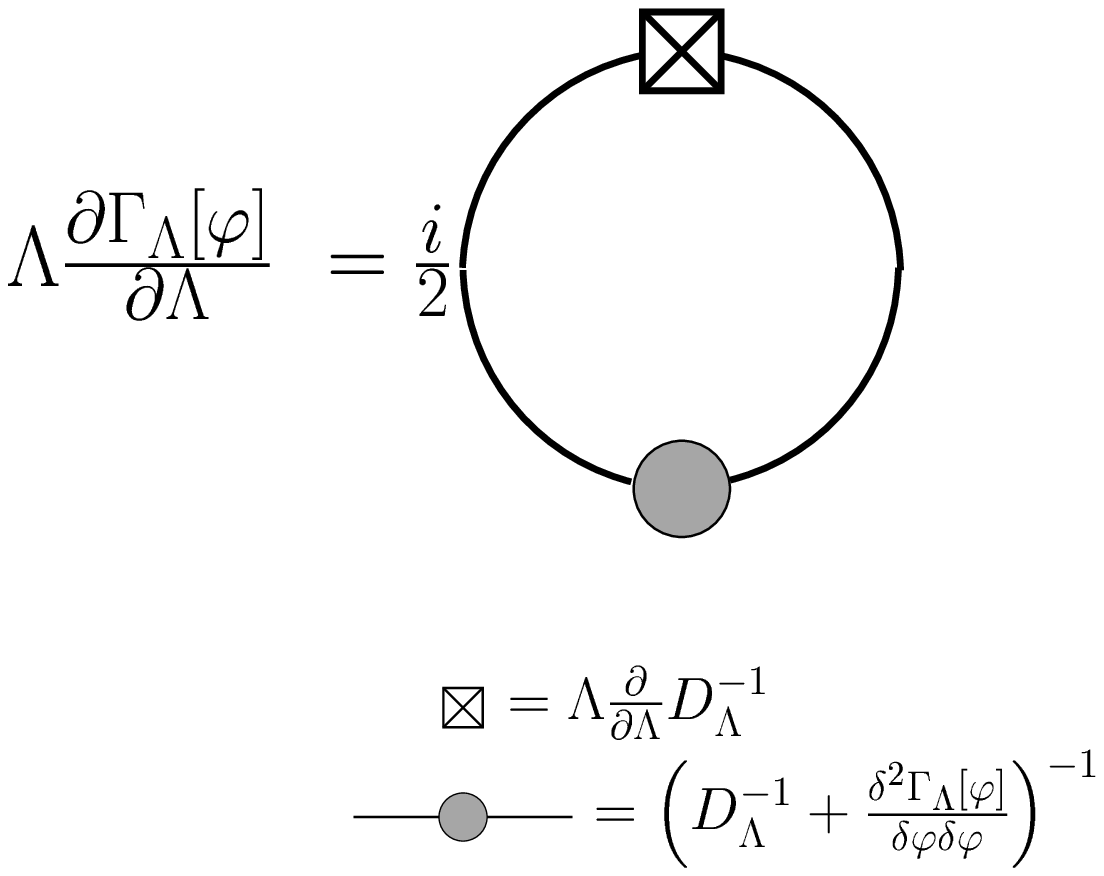,,width=5cm}
\renewcommand{\baselinestretch}{1}\normalsize
{\footnotesize{\footnotesize\caption{Graphical representation of the thermal renormalization
group-equation (\ref{TRG}).}}}
\end{minipage}
\end{center}
\end{figure}
\renewcommand{\baselinestretch}{1.2}\normalsize
Of course, in an interacting theory the full propagator is still a functional
of the fields, making the equation impossible to solve exactly -- nevertheless
the existence of a pictorial description in terms of Feynman-like diagrams
simplifies the discussion of the flow-equations and allows us to easily derive
flow-equations for Green-functions etc.

One should note that in the context of thermal renormalization group-equations
as in the usual real-time formalism one cannot generally dispose the thermal ghost fields
($\varphi_2$ in the above derivation).
Even if one is only interested in Green-functions with physical ($\varphi_1$) external lines,
the facts that equation (\ref{TRG}) is formally of one-loop order and that the
classical vertices do not mix the fields do not help since taking derivatives
of (\ref{TRG}) with respect to $\varphi_1$ introduces {\em{full}} vertices with
external legs. 
These of course will in general mix physical and ghost-fields.
Also note that $\Lambda \partial_\Lambda D_\Lambda^{-1}$ is a non-diagonal
matrix. 
Since the insertion is basically treated like a vertex that mixes physical and
ghost fields in (\ref{TRG}) one gets contributions from all matrix-elements of
the full propagator.

Finally, from the derivation given above it is immediately obvious that the
general method of thermal renormalization group-equations is not restricted to
thermal field theory in equilibrium.
In principle one may use any form of density-matrices for the initial state.
It is clear, however, that if the density-matrix is not bilinear in creation-
and annihilation operators there will be $\Lambda$-dependent contributions also
to the interaction terms in the path-integral (\ref{ZLambda}) and the
renormalization group-equation would not have a simple form.
It is also straightforward to formulate thermal renormalization group-equations
for gauge theories
\cite{DAP2}
or theories involving fermions.

Apart from the fact that the formulation of renormalization
group-equations in the real-time approach to thermal field theories
simplifies the calculation of real-time correlation functions since
one avoids analytical continuation of quantities calculated at
imaginary time arguments, the major difference between the thermal
renormalization group formalism reviewed here and the Matsubara-approach
by means of exact renormalization group-equations is the boundary
condition for $\Lambda \rightarrow \infty$.
As we have seen, the starting value for the effective action in the
TRG-formulation is the full effective action at vanishing temperature.
This should be contrasted to the imaginary-time approach where one
treats quantum- and thermal fluctuations on equal footing. 
There the starting value for the effective action is the renormalized
classical action.
In practice this means that for a computation of finite-temperature
quantities in terms of physical quantities at vanishing temperature a
two-step procedure is necessary
\cite{ChristofNickfiniteT}.
One first has to connect the parameters of the classical action to
observables in a $T=0$-computation\footnote{The exact renormalization
  group-equations imply a specific renormalization scheme. One may
  thus not directly compare results from perturbative computations in,
  say, the $\overline{\mbox{MS}}$-scheme with renormalization group
  results at the same values of the parameters of the renormalized
  action.}
and only afterwards calculate
quantities at finite temperature.
To get control over the relation between
e.g. perturbatively defined $\overline{\mbox{MS}}$-parameters and the
corresponding couplings in the renormalization-scheme defined by the
exact renormalization group approach one has to perform perturbative
calculations also in the RG-approach
\cite{UliSchemes}.
In cases where one needs to go beyond one-loop order, this is an
extremely cumbersome task
\cite{ThomasChristof}.
In this respect the approach used in the present work is very
convenient, since perturbative contributions at vanishing temperature
may be taken into account by using perturbative starting conditions to
any desired order.
This feature implies of course an obvious
limitation to the use of thermal renormalization group-equations in
situations where already the theory at $T=0$ is strongly coupled.
In such a situation one needs input from other nonperturbative methods
such as lattice computations or exact renormalization group-equations
at vanishing temperature.
On the other hand this is a very welcome feature if one wants to
study questions like the validity of perturbative dimensional
reduction, a method that e.g. is at the heart of most of the present
nonperturbative results on the electroweak phase-transition in the early universe
\cite{EWPTreview}
and is also studied in the context of QCD
(for a review see e.g. \cite{DimRed}).
In the case of the electroweak phase-transition there are a number of
lattice studies that do not rely on perturbative dimensional reduction
at least in the bosonic sector of the model
(see for example \cite{EWPT4d} and references therein).
A comparison of these results with studies assuming the validity of
perturbative dimensional reduction indicates that method works fairly
well in this sector
\cite{LatticeDimRed}.
However, to date there exist no nonperturbative checks of the method
in the fermionic sector of the theory which is quantitatively
important due to the large top-Yukawa coupling.
TRG-equations should allow a test of this method also for models with
chiral fermions.

\subsection{Approximation schemes}

Before presenting the results of our study of self-interacting scalar
theories in the next section, we would finally like to briefly discuss some
general strategies to approximately solve equations such as (\ref{TRG})
\cite{ChristofNickCritExp}.
The full effective action $\Gamma_\Lambda$ may be characterized by infinitely
many couplings multiplying the invariants consistent with the symmetries of the
theory under study.
The functional differential equation (\ref{TRG}) is thus equivalent to an
infinite system of coupled nonlinear ordinary differential equations.
Clearly one has to make some approximation to proceed.
All approximations involve a truncation of the system of ordinary differential
equations to another (possibly still infinite) system of equations by
neglecting couplings.

The simplest approximations reduce the number of
couplings considered to a finite set.
A prominent example of such a truncation is the so-called "local polynomial
approximation" (LPA) to the effective potential, supplemented by a derivative
expansion.
This approximation is at the heart of most applications of Wilsonian
RG-equations considered in the literature.
It may be viewed as an expansion in the canonical dimension of the couplings
(see \cite{ChristofNickCritExp}).

Alternatively one may consider truncations that keep an infinite set of
couplings.
This will result in (systems of) partial (integro-)differential equations.
There are two more or less orthogonal ways to proceed:
\begin{enumerate}
\item[(i)] One may perform a derivative expansion of the effective action.
In this case one classifies the invariants by the number of derivatives
appearing.
The coefficients of these invariants are then functions of constant fields.
Considering a scalar theory with one field $\varphi$ and $Z_2$-symmetry for
simplicity, one expands the effective action as
\bea
\Gamma_\Lambda = \int d^d x \left\{ - V_\Lambda(\varphi^2) + \frac{1}{2}
Z_\Lambda(\varphi^2) \left( \partial \varphi \right)^2 + \frac{1}{2}
Y_\Lambda(\varphi^2) \left( u \cdot \partial \varphi \right)^2 + ... \right\}
\label{derexp}
\eea
where we have taken into account the fact that at finite temperature
Lorentz-invariance is broken by the rest-frame of the heat-bath and have
introduced the corresponding four-vector $u_\mu$.
Truncating the effective action at finite order in this expansion, one derives a
set of coupled partial differential equations for the functions
$V_\Lambda(\varphi^2)$, $Z_\Lambda(\varphi^2)$ etc.

This is the approach that we will pursue below. 

\item[(ii)] One may expand the effective action in powers of the
fields,
\bea
\Gamma_\Lambda = \left. \Gamma_\Lambda \right|_{\varphi=0} + \sum_{n=1}^\infty
\frac{1}{n!} \int d^d x_1 ... d^d x_n \gamma_\Lambda^{(n)}(x_1,...,x_n)
\varphi(x_1) ... \varphi(x_n)
\label{gfexp}
\eea
The coefficients of this expansion are the 1PI Green-functions (the full
propagators and vertices) of the theory.
This is the strategy that is naturally followed in the study of Schwinger-Dyson
equations
\cite{SDEreview}.

\end{enumerate}

The approach using the derivative expansion is technically relatively
straightforward and has been used to study scalar theories in three dimensions
\cite{JNC1,CS}
and at finite temperature
\cite{B1,InvSymmBr},
matrix-models
\cite{Juergen},
the abelian as well as an $SU(2)$-Higgs model in three dimensions
\cite{NikosAHM},
and an effective theory for low-energy QCD at finite temperature in the
imaginary-time formalism
\cite{Dirketal}.
It is generally appropriate in situations where the degrees of freedom
considered correspond also to asymptotic states and the anomalous dimensions
are small.
As noted above one obtains a set of coupled partial differential equations for
functions that depend on the invariants that one may construct from the fields
considered (at vanishing momenta).

On the other hand, an expansion in the $n$-point functions rapidly becomes
technically involved with growing $n$.
Considering only the flow-equation for the two-point function of a scalar
theory at $T=0$, one has to deal with a partial integro-differential equation
of the structure
\bea
\Lambda \frac{\partial}{\partial \Lambda} \gamma_\Lambda^{(2)}(p^2) &=&
\int \frac{d^dl}{(2\pi)^d} F_4\left[ \gamma_\Lambda^{(2)}(l^2),
\gamma_\Lambda^{(4)}(p^2,l^2,(lp)) \right] + \nonumber \\ 
&& + \int \frac{d^dl}{(2\pi)^d} F_3\left[
\gamma_\Lambda^{(2)}((l+p)^2),\gamma_\Lambda^{(2)}(l^2),
\gamma_\Lambda^{(3)}(p^2,l^2,(lp)) \right]
\label{feqtpf}
\eea
corresponding to the diagrams depicted in figure 3.
In a theory with vector-fields like QCD the number of invariants that make up a
specific $n$-point function is rapidly growing with $n$.
So far, this approach has been used to study the 2-point functions of pure
Yang-Mills theory
\cite{QCDTzero}
and the four-point function of an effective model for QCD
\cite{UliChristof} (both at vanishing temperature). 

Which of the above mentioned approaches should be considered depends on the
physics of the theory under study.
Generally, the derivative expansion will not be useful in situations where the
momentum-dependence of the correlation functions is strongly
modified by interactions.
\begin{figure}
\begin{center}
\begin{minipage}[b]{.8\linewidth}
\centering\epsfig{file=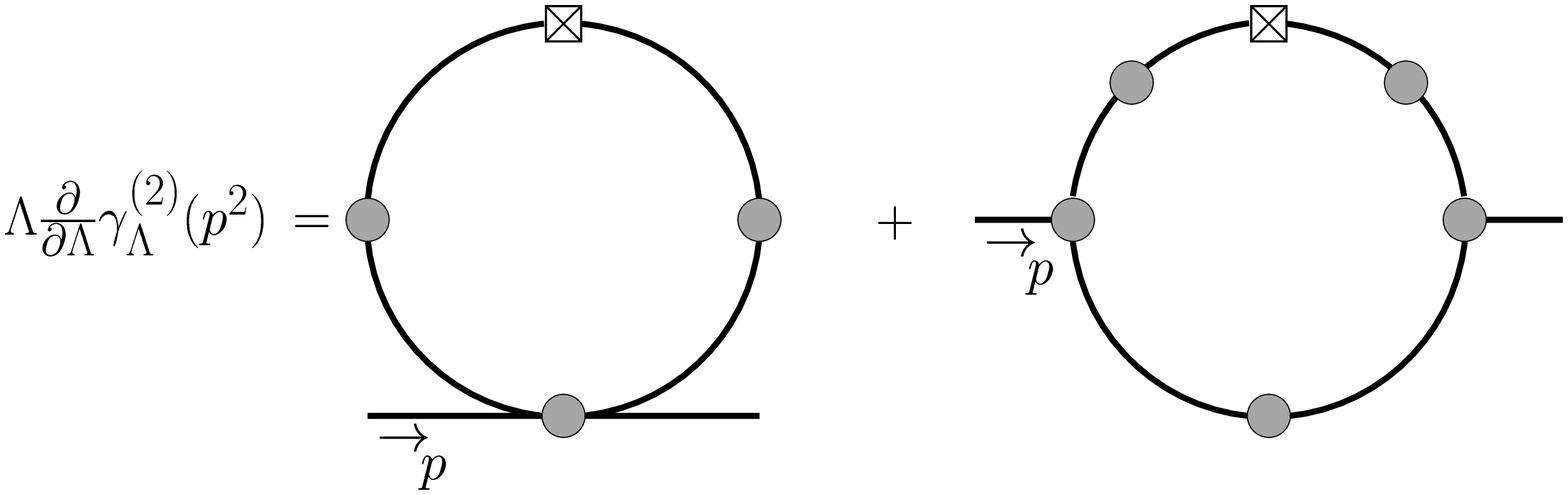,width=9cm}
\renewcommand{\baselinestretch}{1}\normalsize
{\footnotesize{\footnotesize\caption{Exact flow-equation for the two-point
function in a scalar theory.}}}
\end{minipage}
\end{center}
\end{figure}
\renewcommand{\baselinestretch}{1.2}\normalsize

After this more general review we will in the following section consider 
self-interacting scalar theories with an $O(N)$-symmetry at finite
temperature.

\section{Critical behavior of $O(N)$-symmetric scalar field theory}

$O(N)$-symmetric scalar theories are well studied both in 4 dimensions at vanishing temperature
as well as in statistical physics.
The $O(N)$-symmetry of the action may be spontaneously broken at
zero-temperature. 
In this case, one expects symmetry-restoration for large temperatures.
The universal aspects of the corresponding phase-transition are very well known
from considerations of the model in three Euclidean dimensions
(for a review, see 
\cite{ZJ}).
Quantities like the critical exponents are calculated to high orders in the
$\epsilon$-expansion or in perturbation theory at fixed dimension $d=3$, where
in both cases Borel-resummation has to be used to deal with the divergent
series.
There are also results from lattice studies available.

What is less well studied are the non-universal aspects like the critical
temperature as a function of the zero-temperature "physical" couplings.
To study non-universal quantities one really has to consider the model at finite
temperature.
Perturbative studies of the theory at finite temperature are complicated by the
bad infrared behavior near the phase-transition.
Due to this behavior, for $N=1$ even super-daisy resummed perturbation theory
falsely predicts a first-order phase-transition
\cite{PertFirstOrder}.
Alternatively, one may employ an expansion in $1/N$. 
In this case the resummation of daisy-diagrams yields exact results in leading
order in $1/N$
\cite{DolanJackiw} (see also 
\cite{ZJ}).
One finds a second order phase-transition with mean-field values of the
critical exponents $\beta$ and $\eta$, whereas $\nu = 1 + {\mathcal{O}}(1/N)$,
$\delta = 5 + {\mathcal{O}}(1/N)$, $\alpha = -1 + {\mathcal{O}}(1/N)$ and 
$\gamma = 2 + {\mathcal{O}}(1/N)$ (see below for definitions of the critical exponents).

As noted in the introduction, there are a number of methods that correctly
predict a second-order phase-transition of the finite-temperature theory also for $N=1$.
Results for the critical temperature have for example been given in
\cite{ChristofNickfiniteT,Chrisetal1} and we will compare the corresponding values below.

\subsection{Flow-equation for the effective potential}

As stated in the last section, there is no way to solve the TRG-equation
(\ref{TRG}) exactly and we have to make simplifications.
In the theory under consideration here, many studies have shown
that a good truncation consists in performing a derivative expansion of the
effective action and considering only the first few contributions.
This fact is related to the smallness of the anomalous dimensions of the
fields, which govern the scale-dependence of the wave-function
renormalizations.
In this spirit we will in the following approximate the effective action
$\Gamma_\Lambda$ by its value for constant configurations, the effective
potential $V_\Lambda$, and use a standard kinetic term. 
The effective potential is of course still a function of the fields. 
We may at this point use arguments given by
\cite{NS}
to simplify the task of calculating the effective potential considerably.
The line of argument goes as follows:
The free energy of the theory is given by the functional
$\bar{\Gamma}_\Lambda[\varphi]$ which is defined as
\cite{NS}
\bea
\frac{\delta \bar{\Gamma}_\Lambda[\varphi]}{\delta \varphi} =
\left. \frac{\delta \Gamma_\Lambda[\varphi_1, \varphi_2]}{\delta \varphi_1}
\right|_{\varphi_1 = \varphi_2 = \varphi}
\label{Gammabar}
\eea
We are interested in the value of the effective action for
constant field configurations and
define the tadpole $\bar{\Gamma}_\Lambda^{(1)}(\varphi)$ through
\bea
\bar{\Gamma}_\Lambda^{(1)}(\varphi) = 
\left. \frac{\delta \Gamma_\Lambda[\varphi_1, \varphi_2]}{\delta \varphi_1}
\right|_{\varphi_1 = \varphi_2 = \varphi = {\mathrm{const.}}}
\label{tadpole}
\eea
By virtue of the symmetry
\bea
\Gamma_\Lambda[\varphi_1, \varphi_2] = - \Gamma^*_\Lambda[\varphi_2^*,
\varphi_1^*]
\label{symm}
\eea
one may derive relations between $n$-point functions with type-1 and type-2
external legs.

In the following we will discuss $O(N)$-symmetric scalar theories.
We will label the physical- and thermal ghost-fields with subscripts as above,
and use greek letters as superscripts to label the components of the
$O(N)$-vectors ($\alpha, \beta, ... = 1...N$).
We frequently use the $O(N)$-symmetry and write
\bea
\rho = \frac{1}{2} \varphi^\alpha \varphi^\alpha
\label{rho}
\eea

Now consider the renormalization group-equation for the tadpole
(\ref{tadpole}).
Deriving equation (\ref{TRG}) with respect to $\varphi_1^\gamma$ and setting
the fields constant we obtain
\bea
\Lambda \frac{\partial}{\partial \Lambda} \frac{\delta \Gamma_\Lambda}{\delta
\varphi_1^\gamma}(\rho) &=& \frac{i}{2} \int \frac{d^4k}{(2\pi)^4} {\mbox{tr}}
\Biggl\lbrace \left[ - (\Delta_0-\Delta^*_0 )\frac{\Delta^{(\alpha)}_\Lambda
\Delta^{(\alpha)*}_\Lambda}{\Delta_0 \Delta^*_0} \Lambda \frac{\partial
N_f(|\vec{k}|)}{\partial \Lambda} \left( \begin{array}{cc} 1&1\\1&1 \end{array}
\right) 
\right] \delta^{\alpha \beta} \times \nonumber \\
&& \qquad \quad \qquad \times \left( \begin{array}{cc} \frac{\delta^3
\Gamma_\Lambda}{\delta \varphi_1^\beta \delta \varphi_1^\alpha \delta
\varphi_1^\gamma} & \frac{\delta^3
\Gamma_\Lambda}{\delta \varphi_1^\beta \delta \varphi_2^\alpha \delta
\varphi_1^\gamma} \\ \frac{\delta^3
\Gamma_\Lambda}{\delta \varphi_2^\beta \delta \varphi_1^\alpha \delta
\varphi_1^\gamma} & \frac{\delta^3
\Gamma_\Lambda}{\delta \varphi_2^\beta \delta \varphi_2^\alpha \delta
\varphi_1^\gamma} \end{array} \right) \Biggr \rbrace
\label{dLambdatadpole}
\eea
where the trace is over the "thermal" matrix-structure and we have used the
fact that the two-point functions are diagonal with respect to the $O(N)$-indices.
The term in square brackets constitutes the "kernel" of the TRG-equation
\cite{DAP1}
and corresponds to the (modified) logarithmic derivative of the full
(cutoff-)propagator with respect to $\Lambda$ (no sum over $\alpha$),
\bea
\left[ - (\Delta_0-\Delta^*_0 )\frac{\Delta^{(\alpha)}_\Lambda
\Delta^{(\alpha)*}_\Lambda}{\Delta_0 \Delta^*_0} \Lambda \frac{\partial
N_f(|\vec{k}|)}{\partial \Lambda} \left( \begin{array}{cc} 1&1\\1&1 \end{array}
\right) 
\right] &=& \tilde{\partial}_{\ln \Lambda} \left( D_\Lambda^{-1} +
\frac{\delta^2 \Gamma_\Lambda}{\delta \varphi^\alpha \delta \varphi^\alpha} \right)^{-1}
\nonumber \\
&& \hspace*{-6cm} =  - \left( D_\Lambda^{-1} +
\frac{\delta^2 \Gamma_\Lambda}{\delta \varphi^\alpha \delta \varphi^\alpha} \right)^{-1}
\left( \Lambda \frac{\partial}{\partial \Lambda} D_\Lambda^{-1} \right) 
\left( D_\Lambda^{-1} +
\frac{\delta^2 \Gamma_\Lambda}{\delta \varphi^\alpha \delta \varphi^\alpha} \right)^{-1}
\label{kernel}
\eea
where $\tilde{\partial}_{\ln \Lambda}$ denotes a derivative acting only on the
explicit $\Lambda$-dependence of the free propagator $D_\Lambda^{-1}$.

Equation (\ref{dLambdatadpole}) is an obvious generalization of the case of one
scalar field studied in 
\cite{DAP1,B1}.
The trace yields
\bea
{\mbox{tr}} \left( \begin{array}{cc} 1&1\\1&1 \end{array} \right)
\left( \begin{array}{cc} \frac{\delta^3
\Gamma_\Lambda}{\delta \varphi_1^\beta \delta \varphi_1^\alpha \delta
\varphi_1^\gamma} & \frac{\delta^3
\Gamma_\Lambda}{\delta \varphi_1^\beta \delta \varphi_2^\alpha \delta
\varphi_1^\gamma} \\ \frac{\delta^3
\Gamma_\Lambda}{\delta \varphi_2^\beta \delta \varphi_1^\alpha \delta
\varphi_1^\gamma} & \frac{\delta^3
\Gamma_\Lambda}{\delta \varphi_2^\beta \delta \varphi_2^\alpha \delta
\varphi_1^\gamma} \end{array} \right) &=& \sum_{i,j=1}^2 \frac{\delta^3
\Gamma_\Lambda}{\delta \varphi_i^\beta \delta \varphi_j^\alpha \delta
\varphi_1^\gamma} \nonumber \\
&=& \frac{\delta^2}{\delta \varphi^\beta \delta \varphi^\alpha} \frac{\delta
\bar{\Gamma}_\Lambda[\varphi]}{\delta \varphi^\gamma}
\label{thtrace}
\eea

Now consider $i ( \Delta_0 - \Delta_0^*) \frac{\Delta_\Lambda
\Delta_\Lambda^*}{\Delta_0 \Delta_0^*}$. 
The "full" Feynman-propagator $\Delta_\Lambda$ reads
\bea
\Delta^{(\alpha)}_\Lambda = \frac{1}{k^2-m^2+\Pi^{(\alpha)}_\Lambda(k,\varphi) + i \epsilon}
\label{DeltaLambda}
\eea
where the self-energy is given by (no sum over $\alpha$)
\bea
{\mathrm{Re}}\,\, \Pi_\Lambda^{(\alpha)}(k,\varphi) &=& {\mathrm{Re}}
\left. \frac{\delta^2 \Gamma_\Lambda}{\delta \varphi_1^\alpha(-k) \delta
\varphi_1^\alpha (k)}\right|_{\varphi_1=\varphi_2=\varphi={\mathrm{const.}}}
\nonumber \\
{\mathrm{Im}}\,\, \Pi_\Lambda^{(\alpha)}(k,\varphi) &=& \frac{1}{1+2
N_f(|\vec{k}|)} {\mathrm{Im}}
\left. \frac{\delta^2 \Gamma_\Lambda}{\delta \varphi_1^\alpha(-k) \delta
\varphi_1^\alpha (k)}\right|_{\varphi_1=\varphi_2=\varphi={\mathrm{const.}}}
\label{Pi}
\eea
To the considered order in the derivative expansion, the self-energies are
momentum independent and real. 
Then in equation (\ref{dLambdatadpole}) we can replace
\cite{DAP1}
\bea
i\left( \Delta_0 - \Delta^*_0 \right) \frac{\Delta^{(\alpha)}_\Lambda
\Delta^{(\alpha)*}_\Lambda}{\Delta_0 \Delta^*_0} \rightarrow 2 \pi
\delta\left(k^2-m^2+\Pi_\Lambda^{(\alpha)}(\varphi) \right) \Theta \left(
|\vec{k}|^2+m^2-\Pi_\Lambda^{(\alpha)}(\varphi) \right)
\label{Kernel}
\eea
and may use (\ref{symm}) to write (no sum over $\alpha$)
\bea
\Pi_\Lambda^{(\alpha)}(\varphi) = \frac{\delta^2 \Gamma_\Lambda}{\delta
\varphi_1^{\alpha} \delta \varphi_1^{\alpha}} = \frac{\delta^2
\bar{\Gamma}_\Lambda}{\delta \varphi^\alpha \delta \varphi^\alpha}
\label{symm2}
\eea

We finally have to specify the form of the cut-off function $f_\Lambda$.
In order to be able to do the loop-integral in (\ref{dLambdatadpole})
analytically, following 
\cite{DAP1,B1} 
we will choose this function such that 
\bea
N_f(|\vec{k}|) = \frac{1}{e^{\beta \omega_k}-1} \Theta(|\vec{k}|-\Lambda) =
N_{BE}(\omega_k) \Theta(|\vec{k}|-\Lambda)
\label{cutoff}
\eea
where $N_{BE}$ denotes the thermal distribution function for bosons.
In this case we have $\partial_\Lambda N_\Lambda(|\vec{k}|) = N_{BE}(\omega_k)
\delta(|\vec{k}|-\Lambda)$ and the loop integrals are trivial.

We write for the functional $\bar{\Gamma}_\Lambda$
\bea
\bar{\Gamma}_\Lambda[\varphi] = \int d^4 x \left\{ \frac{1}{2} m^2 \varphi^2 -
V_\Lambda(\varphi) + \frac{1}{2} Z_\Lambda(\varphi) \left( \partial \varphi \right)^2
+ \frac{1}{2} Y_\Lambda(\varphi) \left( u \cdot \partial \varphi \right)^2 +
... \right\}
\label{ansatz}
\eea
and, working to lowest order in the derivative expansion, neglect $Z_\Lambda$
and $Y_\Lambda$.
We will allow for spontaneous symmetry breaking along the "1"-direction in the
$O(N)$-group, giving rise to a radial (Higgs-) mode, the field $\varphi^1$, and
$N-1$ Goldstone-modes $\varphi^{\alpha \neq 1}$.
This amounts to evaluating the effective action on a background field
configuration $\varphi^\alpha = \sqrt{2 \rho} \delta^{\alpha 1}+\chi^\alpha(q)$
and putting $\chi^\alpha = 0$ at the end.
We then have (no sum over $\alpha$)
\bea
\Pi_\Lambda^{(\alpha)}(\rho) = m^2 + \frac{\partial^2 V_\Lambda(\rho)}{\partial
\varphi^\alpha \partial \varphi^\alpha}
\label{selfenergy}
\eea
Dropping the subscript $\Lambda$ we find from (\ref{dLambdatadpole})
\bea
\Lambda \frac{\partial}{\partial \Lambda} \frac{\partial
V(\rho)}{\partial \varphi^\gamma} &=&
- \frac{\Lambda^3}{4 \pi^2} \frac{N_{BE}(\omega_1)}{\omega_1} \frac{\partial^3
  V(\rho)}{\partial \varphi^\gamma \partial \varphi^1 \partial
  \varphi^1} \Theta(\omega_1^2) - \nonumber \\
&& - (N-1) \frac{\Lambda^3}{4 \pi^2} \frac{N_{BE}(\omega_2)}{\omega_2}
\frac{\partial^3 V(\rho)}{\partial \varphi^\gamma \partial \varphi^2
\partial \varphi^2} \Theta(\omega_2^2)
\label{feqdVdalpha}
\eea
$\omega_1$ and $\omega_2$ denote the frequencies of the radial
and the Goldstone-modes respectively, i.e.
\bea
\omega_1 = \sqrt{\Lambda^2 + \frac{\partial^2 V}{\partial \varphi^1
\partial \varphi^1}} = \sqrt{\Lambda^2+V'(\rho)+2\rho V''(\rho)} \quad ; \quad \omega_2 = \sqrt{\Lambda^2 +
\frac{\partial^2 V}{\partial \varphi^2 \partial \varphi^2}} = \sqrt{\Lambda^2 +
V'(\rho)}
\label{omegas}
\eea
where primes denote derivatives with respect to $\rho$.
We may finally integrate eq. (\ref{feqdVdalpha}) with respect to
$\varphi^\gamma$ to obtain the flow-equation for the effective potential as 
\bea
\Lambda \frac{\partial}{\partial \Lambda} V(\rho) &=& - T
\frac{\Lambda^3}{2\pi^2} \ln \left[ 1 - \exp\left( - \beta \sqrt{\Lambda^2 +
V'(\rho) + 2 \rho V''(\rho)} \right) \right] \Theta(\Lambda^2 +
V'(\rho) + 2 \rho V''(\rho)) - \nonumber \\
&& - (N-1) T
\frac{\Lambda^3}{2\pi^2} \ln \left[ 1 - \exp\left( - \beta \sqrt{\Lambda^2 +
V'(\rho)} \right) \right] \Theta(\Lambda^2 +
V'(\rho))
\label{feqV}
\eea

We will in the numerical evaluation also use the flow-equation for the minimum
of the potential in the spontaneously broken phase, $\rho_0$.
This is obtained from the condition $\left. V'(\rho) \right|_{\rho_0} =
0$ and reads
\bea
\Lambda \frac{\partial \rho_0}{\partial \Lambda} &=& \frac{\Lambda^3}{4\pi^2}
\left( 3 + 2 \frac{\rho_0 V_0^{(3)}}{V_0''} \right) \frac{1}{\sqrt{\Lambda^2 +
2 \rho_0 V_0''}} \frac{1}{\exp\sqrt{\frac{\Lambda^2}{T^2} + \frac{2 \rho_0
V_0''}{T^2}}-1} \Theta(\Lambda^2+2\rho_0 V_0'') + \nonumber \\
&& + (N-1) \frac{\Lambda^3}{4\pi^2}
\frac{1}{\Lambda}\frac{1}{\exp\sqrt{\frac{\Lambda^2}{T^2}}-1} 
\label{feqrho0}
\eea

As we will see shortly, it is also convenient to define dimensionless
quantities in the following way:
\bea
\lambda = \frac{\Lambda}{T} \quad ; \quad \kappa = \frac{\rho}{\Lambda T} \quad
; \quad u = \frac{V}{\Lambda^3 T} \quad ; \quad \frac{d^i V}{d\rho^i} =
\Lambda^{3-i} T^{1-i} \frac{d^i u}{d\kappa^i}
\label{scaled}
\eea
The corresponding flow-equations read
\bea
\lambda \frac{\partial u}{\partial \lambda} &=& -3 u + u' \kappa - \frac{1}{2
\pi^2} \ln \left[ 1 - \exp\left(- \lambda \sqrt{1+u'+2 \kappa u''} \right)
\right] \Theta(1+u'+2\kappa u'') - \nonumber \\
&& \qquad \quad - \frac{N-1}{2\pi^2} \ln \left[ 1 - \exp\left(- \lambda
\sqrt{1+u'} \right) \right] \Theta(1+u')
\label{fequ}
\eea
and 
\bea
\lambda \frac{\partial \kappa_0}{\partial \lambda} &=& -\kappa_0 +
\frac{\lambda}{4\pi^2} \left( 3 + 2 \frac{\kappa_0 u_0^{(3)}}{u_0''}\right)
\frac{1}{\sqrt{1+2\kappa_0 u_0''}}\frac{1}{e^{\lambda \sqrt{1+2\kappa_0
u_0''}}-1} \Theta(1+2\kappa_0 u_0'') + \nonumber \\
&& \qquad + (N-1) \frac{\lambda}{4\pi^2} \frac{1}{e^\lambda-1} 
\label{feqkappa0}
\eea
As was already pointed out in 
\cite{DAP1,B1},
if one considers the limit where $\Lambda^2 + V' + 2 \rho V''$ and $\Lambda^2 +
V'$ are both $\ll T^2$ one recovers the flow-equations of the purely
three-dimensional theory with a sharp cutoff up to a field independent
contribution.
This is the relevant limit in the scaling-regime close to the phase-transition
and motivates the use of the rescaled variables defined in (\ref{scaled}).
The powers of the temperature appearing in eq.s~(\ref{scaled}) are identical to
those obtained by naive dimensional reduction, whereas the powers of $\Lambda$
are the ones one would use in the purely three-dimensional theory to make the
couplings dimensionless. 

We finally have to give the
starting value for the effective potential in the limit $\Lambda \rightarrow
\infty$.
As discussed in section 2, the effective action in this limit becomes the
effective action of the zero-temperature theory.
Correspondingly the starting value for the effective potential should be the
full effective potential of the theory at vanishing temperature.
At this point we face an obvious problem if we are to use perturbative results
on this quantity.
Consider the effective potential of an $O(N)$-symmetric scalar theory in the
broken phase at $T=0$ as computed e.g. in the $\overline{MS}$-scheme.
Taking the scale to be set by the zero-temperature tree-level Higgs-mass
$m_0^2 = 2 g \rho_0$ to one loop we have
\bea
V_{T=0}(\rho) &=& - g \rho_0 \rho + \frac{g}{2} \rho^2 + \frac{g^2}{64
\pi^2} \left( 3 \rho - \rho_0 \right)^2 \left[ \ln \frac{3\rho-\rho_0}{2\rho_0}
- \frac{3}{2} \right] + \nonumber \\
&& \qquad + (N-1) \frac{g^2}{64\pi^2} \left( \rho - \rho_0 \right)^2 \left[ \ln
\frac{\rho-\rho_0}{2\rho_0} - \frac{3}{2} \right]
\label{V0}
\eea
On the right hand side of the flow-equation (\ref{feqV}) we need the first two
derivatives of the potential with respect to $\rho$ at arbitrary values of the
field.
Now from equation (\ref{V0}) it is obvious that the second derivative of the
potential has logarithmic singularities at $\rho = \frac{\rho_0}{3}$ and, for
$N>1$, at $\rho = \rho_0$.
This is of course a well known fact which for example forces us to define
the renormalized quartic coupling in the massless theory with $N=1$ away from
the point $\rho = 0$
\cite{ColemanWeinberg}.
As long as we work with a polynomial expansion,
i.e. as long as we choose to write the potential as
\bea
V(\rho) = V(\tilde{\rho}) + \sum \frac{V^{(n)}}{n!} (\rho - \tilde{\rho})^n
\label{LPA}
\eea
and derive flow-equations for the couplings $V^{(n)}$, this is no
problem.
For $N=1$ we may use $\tilde{\rho}=\rho_0$, for $N>1$ one may
still proceed as long as one does not expand the potential about its minimum.
Even in that case, we could in principle avoid any infrared problems by using
the perturbative effective potential evaluated with an infrared cut-off in the
loop integration.
This would compare closely to the approach followed in 
\cite{ChristofNickfiniteT},
where a local polynomial approximation was used for the potential and the
zero-temperature renormalized couplings were defined at some non-vanishing
scale.
In this work we will restrict ourselves to small values of the
zero-temperature couplings and neglect the quantum-corrections to the effective
potential, that is we use as a starting value for the integration of the
flow-equation (\ref{feqV}) the tree-level potential
\bea
V_{\Lambda_0}(\rho) = - g \rho_0 \rho + \frac{g}{2} \rho^2
\label{VLambda0}
\eea
This simplification has of course no effect on the universal aspects of the
critical behavior such as the critical exponents, certain ratios of critical
amplitudes or the scaling equation of state.
For non-universal quantities such as the critical temperature we will have
corrections from the (logarithmic) four-dimensional running of the couplings.
These corrections should however be small for the values of the couplings
considered in the present work.
We now discuss the results of a solution of the flow-equations given in
(\ref{feqV}), (\ref{feqrho0}),  (\ref{fequ}) and (\ref{feqkappa0})
with the starting value given by (\ref{VLambda0}) (or the corresponding scaled
potential according to (\ref{scaled})) for different values of $N$.
For the numerical work, the methods proposed in 
\cite{matching} 
are used.

\subsection{Numerical results}

We start by discussing the results for the case of one scalar field,
corresponding to the Ising model of statistical mechanics.
The first question concerns the order of the phase-transition.
As was already pointed out above, daisy-resummed perturbation theory fails in
predicting the correct critical behavior of the theory in this case.
In the framework of thermal renormalization group-equations, the case $N=1$ has
already been addressed in 
\cite{DAP1,B1},
where the transition was found to be second order and critical exponents have
been given.

We display the dependence of the minimum of the potential on the external scale
$\Lambda$ at different temperatures in figure 4.
\begin{figure}
\begin{center}
\begin{minipage}[b]{.8\linewidth}
\centering\epsfig{file=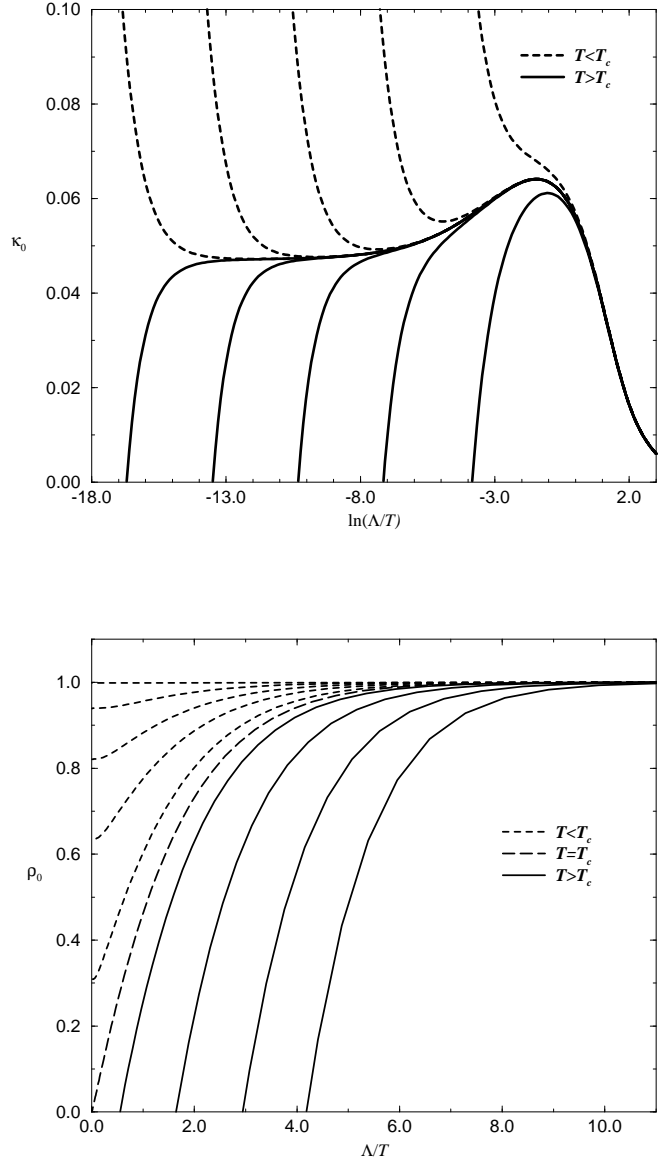}
\renewcommand{\baselinestretch}{1}\normalsize
{\footnotesize{\footnotesize\caption{Upper panel: The dimensionless minimum as
a function of $\ln{\Lambda/T}$ for various values of the temperature around
$T_c$. For temperatures above $T_c$ the minimum vanishes at some finite scale
$\Lambda_{\mathrm{symm}}$, for $T<T_c$ the dimensionless minimum diverges as
$\Lambda \rightarrow 0$.
Lower panel: The dimensionful minimum $\rho_0$ (in units of $\rho_0(T=0)$) as a function of $\Lambda/T$.
For temperatures below the critical temperature the dimensionful minimum
approaches a constant value that is a decreasing function of the temperature
until it vanishes at $T_c$ as $\Lambda \rightarrow 0$.
For low temperatures thermal fluctuations play
effectively no r{\^o}le and $\rho_0$ almost remains at the starting value.}}}
\end{minipage}
\end{center}
\end{figure}
\renewcommand{\baselinestretch}{1.2}\normalsize
In the upper panel, the dimensionless minimum $\kappa_0$ is displayed as a
function of $\ln \frac{\Lambda}{T}$ for several temperatures around the
critical one (we have chosen $g=0.1$ here).
We start the evolution of the potential at some large value of $\Lambda/T$ in
the broken phase, where $\rho_0(T=0)$ is used to set the scale and all
dimensionful quantities are given in units of $\rho_0(T=0)$.
For temperatures larger than the critical one (solid lines), the minimum
vanishes at some nonzero scale $\Lambda_{\mathrm{symm}}$ and the theory is in
the symmetric phase.
On the other hand, for temperatures below $T_c$ the dimensionful minimum
$\rho_0$ approaches some constant as $\Lambda \rightarrow 0$ (this is
seen in the lower panel of fig.~4, where we display the dimensionful minimum as
a function of $\Lambda/T$).
According to (\ref{scaled}) $\kappa_0$ then diverges.
We see from figure 4 that at the critical temperature, the dimensionless
minimum asymptotically reaches a finite nonvanishing value, $\kappa^*_0$.
As is expected from dimensional reduction and was explicitly demonstrated in 
\cite{B1},
this value is the fixed-point value of the corresponding three-dimensional
theory.
The phase-transition is of second order and the theory is in the universality
class of the three-dimensional Ising model.
Before turning to the universal behavior of the theory, let us briefly discuss
the critical temperature as a function of the zero-temperature coupling.
Our results on this quantity are displayed in figure 5, where we plot
the ratio of $T_c$ and the naive perturbative result, given by
\bea
T_{c,1-{\mathrm{loop}}} = \sqrt{\frac{24}{N+2}} \sqrt{\rho_0}
\label{pertTc}
\eea
as solid line and compare to the values found in the ERG-approach in the
Matsubara-formalism in 
\cite{ChristofNickfiniteT}
as squares.
We find good agreement as expected\footnote{Note that the results quoted in 
\cite{ChristofNickfiniteT}
were obtained with an exponential cutoff-function rather than with the
$\Theta$-function cutoff used here.}.
\begin{figure}
\begin{center}
\begin{minipage}[b]{.8\linewidth}
\centering\epsfig{file=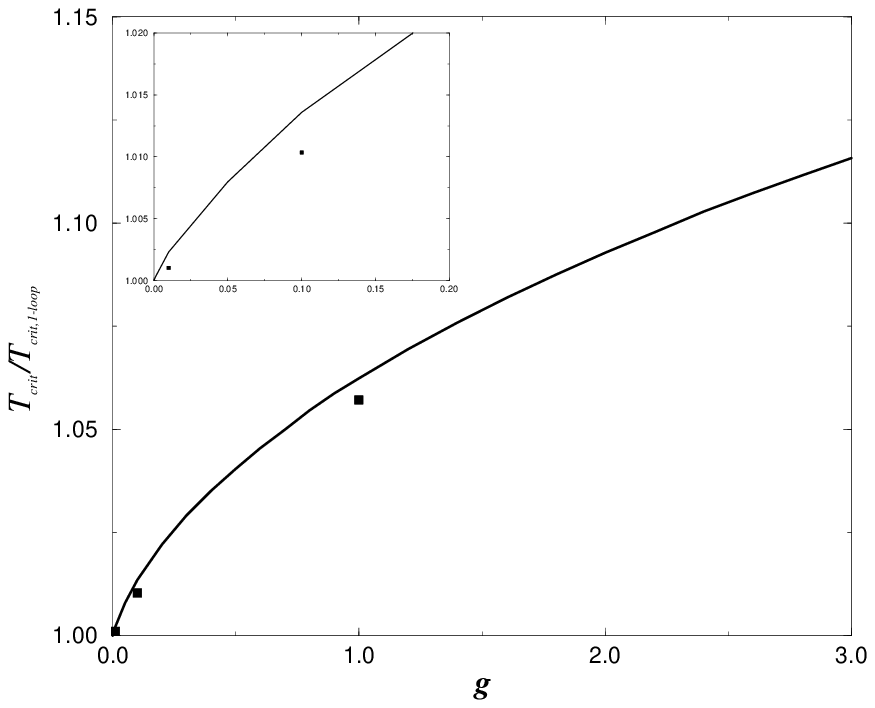}
\renewcommand{\baselinestretch}{1}\normalsize
{\footnotesize{\footnotesize\caption{The ratio $T_c/T_{c,1-{\mathrm{loop}}}$
as a function of the zero-temperature coupling $g$ for $N=1$.
For comparison results from 
\cite{ChristofNickfiniteT}
are also given (squares).}}}
\end{minipage}
\end{center}
\end{figure}
\renewcommand{\baselinestretch}{1.2}\normalsize

We finally discuss the results on the universal behavior as obtained
from the TRG.
The universal critical behavior of a given field theory is summarized in a
number of critical exponents, certain ratios of critical amplitudes, and the
scaling equation of state (see e.g.~\cite{ZJ}).
We will start by discussing critical exponents and amplitudes.
These quantities encode the behavior of different observables as the temperature
approaches its critical value.
Specifically, we will consider the following quantities:
\begin{enumerate}
\item[(i)] The (unrenormalized) mass of the order-parameter field (in the
language of statistical mechanics the inverse magnetic susceptibility in zero field).
The susceptibility diverges at the phase-transition, corresponding to a
behavior of the mass according to
\bea
M_\pm^2 = \left( C_\pm \right)^{-1} \left| \frac{T_c-T}{T_c} \right|^\gamma
\label{Cgamma}
\eea
This defines the critical exponent $\gamma$ and the amplitudes $C_\pm$. 
The ratio $C_+/C_-$ is a universal quantity.
\item[(ii)] The quartic coupling $g(\Lambda \rightarrow 0)$ goes to zero at the
critical temperature according to 
\bea
g_\pm = \ell_\pm \left| \frac{T_c-T}{T_c} \right|^\nu
\label{ellnu}
\eea
The exponent $\nu$ also governs the behavior of the renormalized mass at the
critical temperature.
Since we have approximated the wave-function renormalizations to be constant,
$\nu$ should be equal to $\gamma/2$.
\item[(iii)] The dimensionful minimum of the effective potential (corresponding
to the spontaneous magnetization) approaches 0 as the
temperature approaches $T_c$ from below, where
\bea
\varphi_0 = B \left(\frac{T_c-T}{T_c}\right)^\beta
\label{Bbeta}
\eea
\item[(iv)] Finally, the behavior of the effective potential at the critical
temperature is described for small values of $\rho$ by (remember $\varphi =
\sqrt{2 \rho}$)
\bea
\frac{\partial V}{\partial \varphi} = D \varphi^\delta
\label{delta}
\eea
\end{enumerate}

We stress that we are considering the theory explicitly at finite temperature.
Thus one may expect the scaling behavior encoded in the critical exponents
only in the region where dimensional reduction is effective. 
The relation (\ref{delta}) will for example not hold for large $\varphi \grgl
T$ where the three-dimensional limit is not reached.
Effective exponents taking into account this "dimensional crossover" are
discussed for example in 
\cite{ChrisReview} and the second reference of \cite{Mike1}.
Our results for the critical exponents defined above are given in table 1,
the results for some critical amplitude-ratios and critical couplings are
displayed in table 2.
In both tables, we give results obtained by other approaches for
comparison.
Let us make some comments regarding the results presented in the tables.
We should first note that the exponents may, apart from the possibility
discussed above, also be determined from the scaling equation of state to be
discussed below.
We have in particular obtained the value of $\delta$ in this way. 
The value is completely consistent with $\delta=5$ which was found using the
thermal renormalization group, but extracting $\delta$ according to the above
definition, eq. (\ref{delta}) in 
\cite{B1}.
In fact, one can show that $\delta=5$ is the exact solution to the fixed-point
equations in the three-dimensional theory in lowest order in the derivative
expansion
\cite{Morrisdelta}.

\renewcommand{\baselinestretch}{1}\normalsize
\begin{table}
\begin{center}
\begin{tabular}{||c||c|c|c|c|c||}
\hline
 & $\beta$ & $\gamma$ & $\delta$ & $\eta$ & $\nu$ \\
\hline
 This work & 0.345 & 1.37 & 4.97 & -- & 0.67  \\
 TRG+LPA \cite{DAP1} & & & 3.57 & 0.015 & 0.58 \\
\hline
3d ERG + LPA \cite{ChristofNickCritExp} & 0.333 & 1.247 & & 0.045 & 0.638 \\
3d ERG \cite{JNC1} & 0.336 & 1.258 & 4.75 & 0.044 & 0.643 \\
3d ERG + WFR \cite{CS} & 0.330 & 1.232 & & 0.047 & 0.631 \\
\hline
Best values \cite{ZJ} & 0.325 & 1.240 & 4.81 & 0.032 & 0.630 \\
\hline
\end{tabular}
\caption{Critical exponents from different approaches ($N=1$)}
\end{center}
\end{table}
\renewcommand{\baselinestretch}{1.2}\normalsize

The other exponents given in the table have been determined in a number of
ways:
We have determined $\beta$ from (\ref{Bbeta}) above and from the scaling
equation of state and find values of $0.3458$ and $0.3451$, identical within
the numerical accuracy.
$\gamma$ was obtained from the definition given in (\ref{Cgamma}) and from the
asymptotic behavior of the Widom scaling function parametrizing the equation of
state and we find $1.369$ from the
EOS, $1.368$ from (\ref{Cgamma}) if $T_c$ is approached from above and $1.369$
if we approach the critical temperature from below.
Finally $\nu$ is obtained from (\ref{ellnu}) to be $0.679$ coming from $T>T_c$
and $0.663$ for $T \rightarrow T_c$ from below.
All in all the critical exponents are rather robust with respect to details of
the numerical procedure.
The deviation from the best values given also in table 1 is due to the
approximations made in reducing the exact renormalization group-equation to the
flow-equation for the effective potential, eq.~(\ref{feqV}), namely the neglect
of higher orders in the derivative expansion.
This also prevents us from studying the critical exponent $\eta$ which is
defined as the anomalous dimension at the critical temperature as $\Lambda
\rightarrow 0$.
From the table it is also clear that we could improve on the results for the
exponents and amplitudes by taking into account the wave-function
renormalizations, as has been done in the work listed as ERG in table 1.
In \cite{ChristofNickCritExp} the potential was expanded around the minimum and
only a finite number of couplings was kept (LPA). 
On the other hand, the authors went beyond leading order in the derivative
expansion.
The LPA was dropped in \cite{JNC1}, where however the wave-function
renormalization was still assumed to be field independent.
This approximation is finally also given up in \cite{CS}.
We should keep in mind that all these results were obtained in the
framework of a three-dimensional effective theory rather than in the full
temperature-dependent problem.

\renewcommand{\baselinestretch}{1}\normalsize
\begin{table}
\begin{center}
\begin{tabular}{||c||c|c|c|c||c|c|c||}
\hline
 & $C_+/C_-$ & $R_\chi$ & $\lambda_{4*}$ & $\lambda_{6*}$ & $B$ & $C_+$ & $D$
 \\ 
\cline{6-8}
 & & & & & \multicolumn{3}{c||} {$(g_{T=0} = 0.1)$} \\ 
\hline
 This work & 4.50 & 1.76 & 19.6 & 1725  & 1.05 & 0.54 & 2.73\\
\hline
3d ERG + LPA \cite{ChristofNickCritExp} & & & 27.8 & 1311 & & &\\
3d ERG \cite{JNC1} &  4.29 & 1.61 &  &  & & &\\
3d ERG + WFR \cite{CS} & 4.966 & 1.647 &  &  & & &\\
\hline
Monte-Carlo \cite{Tsypin2} & & & 23.3 & 1476 & & &\\
\hline
Best values \cite{ZJ} &  4.95 & 1.65 &  &  & & & \\
\hline
\end{tabular}
\caption{Critical amplitude ratios and couplings from different approaches
($N=1$). The last three entries are non-universal.}
\end{center}
\end{table}
\renewcommand{\baselinestretch}{1.2}\normalsize

Let us now discuss the results presented in table 2.
Here we give the universal ratios of critical amplitudes $C_+/C_-$ and $R_\chi$
as well as the universal critical couplings $\lambda_{4*}$ and $\lambda_{6*}$
and the non-universal amplitudes $B$, $C_+$ and $D$ for $g_{T=0} = 0.1$.
The definition of $C_\pm$ has been given in (\ref{Cgamma}). 
$R_\chi$ is a universal combination of several non-universal quantities and
reads
\bea
R_\chi = C_+ D B^{\delta-1}
\label{Rchi}
\eea
We have extracted the numbers for the amplitudes used here from the equation of
state to be discussed below.
We have also extracted the ratio $C_+/C_-$ from the behavior of the
unrenormalized mass as given in (\ref{Cgamma}). 
In this case we find $C_+/C_- = 4.58$, consistent with the value
quoted in the table.
As noted in the discussion of the critical exponents, also in the case of the
amplitudes the major source of error is the derivative expansion used here,
as may be seen by again comparing with the results given in 
\cite{CS}.

The universal critical couplings given in the table are defined by
\bea
V(\varphi) = \sum_{n=2}^\infty \frac{\lambda_n}{n!} \varphi^n
\label{critcoup}
\eea
and are taken at the critical temperature, i.e. in the scaling limit.
$\lambda_{4*}$ and $\lambda_{6*}$ are thus simply related to the second and
third derivative of the potential at the critical temperature with respect to
$\rho$.
The values given for comparison have been found through the ERG, using a local
polynomial expansion and on the lattice, in both cases working in three
dimensions.

Let us then turn to the scaling equation of state.
In the language of statistical mechanics the equation of state relates the
temperature, the order parameter and the external field.
In the present setting, the external field $H$ is given by $H =
V'(\rho) \varphi$ and the equation of state has the following scaling
form:
\bea
\frac{\partial V}{\partial \rho} = \left( \sqrt{2 \rho} \right)^{\delta-1} f(x) \qquad ;
\qquad x = \frac{T-T_c}{T_c}\left(\sqrt{2 \rho}\right)^{-1/\beta}
\label{EOS}
\eea
The function $f(x)$ is known as the Widom scaling function and is universal up
to normalization and a rescaling of $x$.
It encodes information about the universal critical behavior and several
amplitudes and critical exponents may be obtained from $f$ in different limits.
For the amplitudes and exponents treated above we have the following relations
involving the equation of state:
The value of $f$ at $x=0$ gives the amplitude $D$,
\bea
\lim_{x \rightarrow 0} f(x) = D 
\label{D}
\eea
The asymptotic behavior of $f$ in the symmetric phase is
\bea
\lim_{x \rightarrow +\infty} f(x) = \left(C_+\right)^{-1} x^\gamma
\label{Cgamma2}
\eea
The amplitude $B$ and the critical exponent $\beta$ are related to the zero of
$f$ in the broken phase through
\bea
f\left(-B^{-1/\beta}\right) = 0
\label{Bbeta2}
\eea
and the amplitude $C_-$ is obtained from the zero of $f$ and the value of the
derivative at the zero through
\bea
\left( C_- \right)^{-1} = \frac{1}{\beta} f'\left(-B^{-1/\beta}\right)
B^{\delta-1-\frac{1}{\beta}}
\label{Cm}
\eea

We have pointed out that the normalization of $f$ and of $x$ are non-universal.
We will in the following figures compare our results with results from exact
renormalization group-equations in three dimensions 
\cite{JNC1}
and with results from Monte-Carlo simulations in the broken 
\cite{Tsypin}
and in the symmetric phase
\cite{Tsypin2} (figure 6).
\begin{figure}[t]
\begin{center}
\begin{minipage}[b]{.8\linewidth}
\centering\epsfig{file=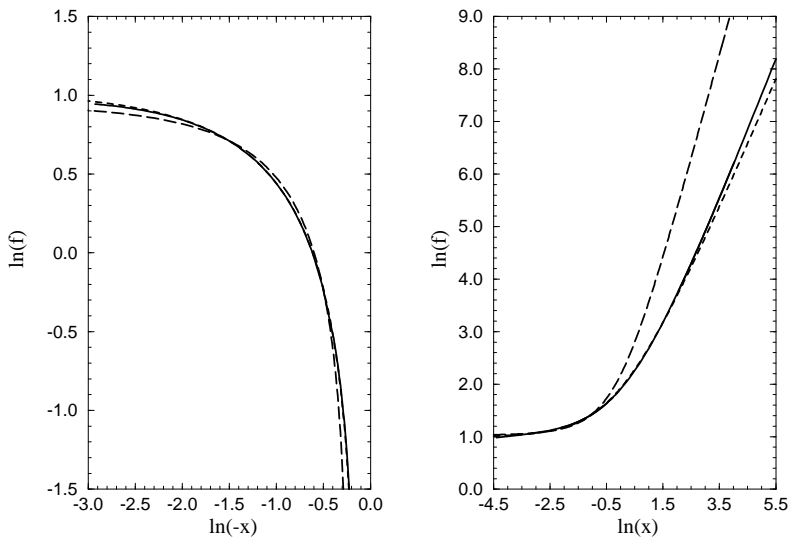}
\renewcommand{\baselinestretch}{1}\normalsize
{\footnotesize{\footnotesize\caption{The logarithm of the Widom scaling
function $f(x)$ for $N=1$ (solid lines). The left panel shows $f(x)$ in the broken phase,
corresponding to $x<0$ whereas the right panel gives $f$ in the symmetric
phase.
The short dashed lines are from the ERG in three dimensions
\cite{JNC1},
the long dashed lines give results from lattice simulations (broken phase: 
\cite{Tsypin}, symmetric phase: \cite{Tsypin2}). 
}}}
\end{minipage}
\end{center}
\end{figure}
\renewcommand{\baselinestretch}{1.2}\normalsize
In order to do so, we fix the normalization by demanding equality of the
results given in form of a numerical fit in 
\cite{JNC1}
and approximate polynomial expressions in 
\cite{Tsypin,Tsypin2}
with our results for 2 values of $x$.
Since the lattice simulations used for the comparison were obtained on
different lattices we have used different rescalings for the Monte-Carlo
results in the left and right
panels in figure 6, corresponding to the broken and symmetric phase
respectively.
In order to have a fair comparison also of the results from 
\cite{JNC1} with the lattice results in both phases, we have also rescaled
these results differently in both panels.

In principle, for a comparison of our results with the ones of 
\cite{JNC1} 
we only have two free parameters corresponding to two values of $x$ where
equality may be imposed.
Thus we also present in figure 7 a comparison of our curve with the curve given
by 
\cite{JNC1}
for different normalizations:
The solid line in figure 7 corresponds to our result in the symmetric phase,
the short dashed line is the same as in the right panel of figure 6, obtained
by equating the results at two values of $x$ in the symmetric phase, whereas
the dashed-dotted line is the result from 
\cite{JNC1} 
if we use the two values of $x$ from the broken phase (left panel of fig.~6)
for normalization for the whole range of $x$.
\begin{figure}[t]
\begin{center}
\begin{minipage}[b]{.8\linewidth}
\centering\epsfig{file=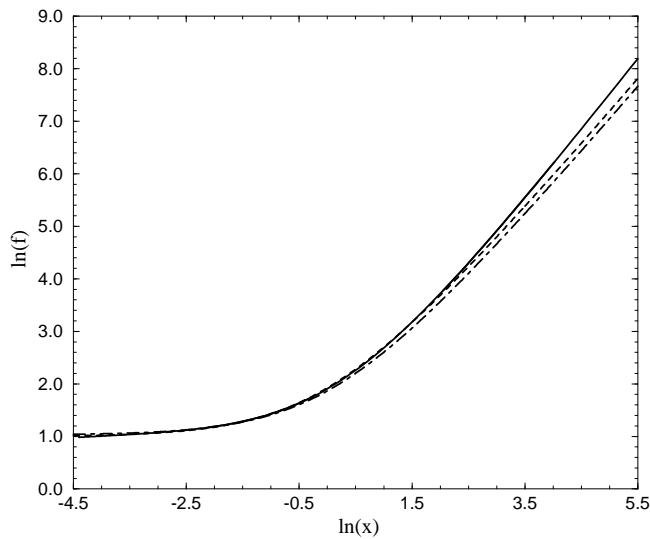}
\renewcommand{\baselinestretch}{1}\normalsize
{\footnotesize{\footnotesize\caption{$f(x)$ for $N=1$ in the symmetric phase
compared to the three-dimensional ERG results from
\cite{JNC1} for different normalizations (see text for details).
}}}
\end{minipage}
\end{center}
\end{figure}
\renewcommand{\baselinestretch}{1.2}\normalsize

The curves obtained in this work and the results given in 
\cite{JNC1} are almost identical in the broken phase (left panel of fig.~6) and show only
a slight deviation in the symmetric phase for large $x$, connected to the
different results for the critical exponent $\gamma$ (see table 1).
The agreement with the results from lattice simulations is
rather satisfactory, as was already noted in
\cite{B1}
for the broken phase.
In view of the sizable deviations of results obtained in the framework of the
$\epsilon$-expansion or other perturbative approaches from the lattice
results\footnote{For a comparison of the lattice results with results
  of the $\epsilon$-expansion, see 
\cite{Tsypin,B1}.},
this is a nontrivial achievement.
The deviation in the asymptotic behavior for large positive $x$ in
figure 6 is of no concern. 
In this region the approximate polynomial expression given in 
\cite{Tsypin2} is incompatible with the known behavior
corresponding to (\ref{Cgamma2}) -- it would give an exponent $\gamma=2$ in
contrast to the known results (table 1).

After presenting detailed results for $N=1$, we now proceed to a study of the
case $N=4$.
This case is interesting, since the is expected to govern the universal
behavior of the two-flavor chiral phase-transition in the chiral limit.
The arguments in favor of this conjecture rely on dimensional reduction and
the fact that in the imaginary-time formulation the fermionic degrees of
freedom obey anti-periodic boundary conditions and thus have no static modes.
The infrared behavior should then be dictated by the bosonic excitations
alone, which in this case are the pions and the "sigma"
(for a more complete discussion of the chiral phase-transition in the
two-flavor case also away from the chiral limit in the imaginary-time formalism
see e.g. 
\cite{Dirketal}).
We should also mention that due to the existence of massless degrees of
freedom also in the broken phase away from the critical temperature
perturbative calculations for $N>1$ are even more infrared problematic then for
the case discussed above.

We start our discussion by plotting the critical temperature as a
function of the zero-temperature coupling $g(T=0)$ in figure 8 (solid line).
We again display for comparison the values obtained by Tetradis and Wetterich
\cite{ChristofNickfiniteT}.
As for the case $N=1$ we find excellent agreement of the results.
\begin{figure}
\begin{center}
\begin{minipage}[b]{.8\linewidth}
\centering\epsfig{file=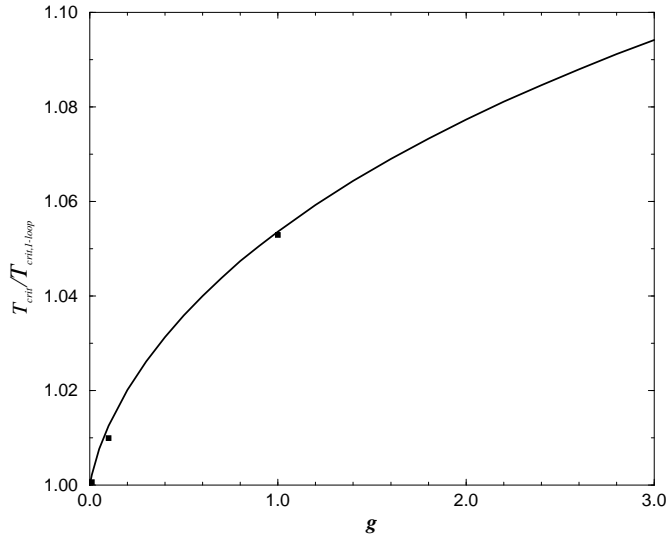}
\renewcommand{\baselinestretch}{1}\normalsize
{\footnotesize{\footnotesize\caption{The ratio $T_c/T_{c,1-{\mathrm{loop}}}$
as a function of the zero-temperature coupling $g$ for $N=4$.
For comparison results from 
\cite{ChristofNickfiniteT}
are also given.}}}
\end{minipage}
\end{center}
\end{figure}
\renewcommand{\baselinestretch}{1.2}\normalsize

Next we turn to the critical equation of state.
This is displayed in figure 9, where as for $N=1$ we display $\ln(f)$ as a
function of $\ln(x)$ in the symmetric phase ($x>0$) in the right panel and
as a function of $\ln(-x)$ in the broken phase ($x<0$) in the left panel.
As discussed above one obtains critical exponents and amplitudes from the
scaling function, and we collect the values together with some results from
other approaches in tables 3 and 4.

\begin{figure}
\begin{center}
\begin{minipage}[b]{.8\linewidth}
\centering\epsfig{file=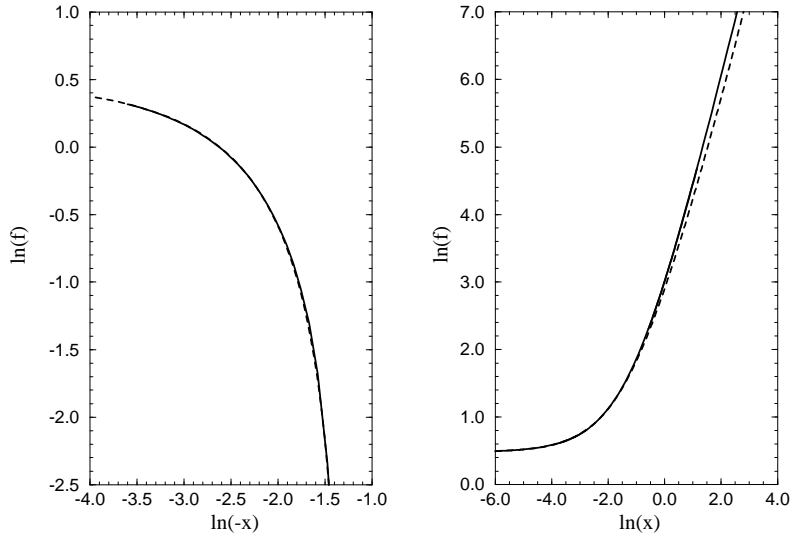}
\renewcommand{\baselinestretch}{1}\normalsize
{\footnotesize{\footnotesize\caption{The Widom scaling function for
      $N=4$ (solid line) in the broken (left panel) and the symmetric
      (right panel) phase. For comparison we give the suitably
      normalized results from \cite{JNC2} (dashed line).
}}}
\end{minipage}
\end{center}
\end{figure}
\renewcommand{\baselinestretch}{1.2}\normalsize
Considering the Widom scaling function, we again observe from figure 9
almost perfect agreement of our result (the solid curve) with results
from the ERG given in 
\cite{JNC2} (dashed curve) in the broken phase.
In the symmetric phase, there is a slight deviation for large $x$,
which again is due to the different values found for the exponent
$\gamma$ (table 3).
In figure 9, we have normalized the results from 
\cite{JNC2}
to our results for two positive values of $x$ close to $x=0$.
A comparison of the results from the ERG and results from other
approaches including lattice and $\epsilon$-expansion may be found in
\cite{JNC2} and again shows that the lattice results on the equation
of state compare well with results from the Wilson renormalization
group, whereas the $\epsilon$-expansion differs from the Monte-Carlo
results.

\renewcommand{\baselinestretch}{1}\normalsize
\begin{table}
\begin{center}
\begin{tabular}{||c||c|c|c|c|c||}
\hline
 & $\beta$ & $\gamma$ & $\delta$ & $\eta$ & $\nu$ \\
\hline
 This work & 0.433 & 1.73 & 5.0 & -- & 0.86  \\
\hline
3d ERG + LPA \cite{ChristofNickCritExp} & 0.409 & 1.556 & & 0.034 & 0.791 \\
ITF ERG \cite{JNC2} & 0.407 & 1.548 & 4.80 & 0.0344 & 0.787 \\
\hline
3d PT \cite{Bakeretal} & 0.38 & 1.44 & 4.82 & 0.03 & 0.73 \\
3d MC \cite{KanayaKaya} & 0.384 & 1.48 & 4.85 & 0.025 & 0.748 \\
\hline
\end{tabular}
\caption{Critical exponents from different approaches ($N=4$)}
\end{center}
\end{table}
\renewcommand{\baselinestretch}{1.2}\normalsize
\renewcommand{\baselinestretch}{1}\normalsize
\begin{table}
\begin{center}
\begin{tabular}{||c||c|c|c|c||c|c|c||}
\hline
 & $C_+/C_-$ & $R_\chi$ & $\lambda_{4*}$ & $\lambda_{6*}$ & $B$ & $C_+$ & $D$
 \\ 
\cline{6-8}
 & & & & & \multicolumn{3}{c||} {$(g_{T=0} = 0.1)$} \\ 
\hline
 This work & -- & 1.38 & 9.4 & 475 & 1.81 & 0.0797 & 1.62 \\
\hline
ITF ERG \cite{JNC2} &  -- & 1.02 &   &  & & & \\
\hline
\end{tabular}
\caption{Critical amplitude ratios and couplings from different approaches
($N=4$). The last three entries are non-universal.}
\end{center}
\end{table}
\renewcommand{\baselinestretch}{1.2}\normalsize
For the critical exponents given in table 3 we note a slightly larger
deviation of our values from those obtained from the lattice 
\cite{KanayaKaya}
or from higher order perturbation theory
\cite{Bakeretal}.
Although the values obtained from the ERG in different approximations
also show a larger deviation than was obtained for $N=1$, in the case
$N=4$ our results differ also from the ERG-results by up to $10 \%$.
There are a number of reasons for this deviation.
First, as in the case of $N=1$ the derivative expansion induces
an error in the critical exponents.
We expect the error from the neglect of the wave-function
renormalization $Z_\Lambda$ to be responsible for most of the
difference between the values found in the present work and the ones
given in 
\cite{ChristofNickCritExp} and 
\cite{JNC2}.
We have also checked in the three-dimensional effective theory and
using the local polynomial expansion for the effective action how much
the choice of a $\Theta$-function cutoff in (\ref{cutoff}) affects the
result (in lowest order in the derivative expansion).
In principle, the choice of the cutoff-function has no effect on the
results if one solves the full RG-equation.
Since we have to make approximations, a slight cutoff-dependence is
expected.
In the case we have checked, the results for the critical exponents
using a sharp cutoff and 
using an exponential cutoff as usually used in the framework of the
ERG differ by about $3 \%$.

There is however an additional contribution in first order in the
derivative expansion for $N>1$, being of the structure
$\tilde{Z}_\Lambda(\rho) \varphi^\alpha \varphi^\alpha \partial^2
\varphi^\beta \varphi^\beta$.
This term should account for much of the remaining difference to the
exact exponents.

The same remarks apply to the universal amplitude-ratio $R_\chi$ which
is compared to the value found in 
\cite{JNC2}
in table 4.
In view of these uncertainties the values for the exponents and
amplitudes found in
the present work are consistent with the expectations, although they
certainly leave room for improvement along the lines discussed above.

Finally, we consider the $N$-dependence of the critical temperature to compare
with the leading order result of the large-$N$ expansion.
In figure 10 we have plotted our values for $g(T=0) = 0.1$ as a function of $N$,
where again the critical temperatures are given in units of $\sqrt{\rho_0}$
(solid line).
\begin{figure}
\begin{center}
\begin{minipage}[b]{.8\linewidth}
\centering\epsfig{file=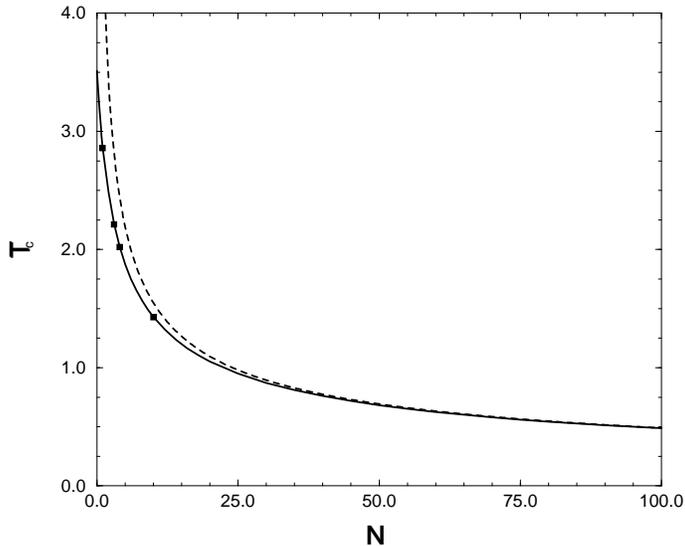}
\renewcommand{\baselinestretch}{1}\normalsize
{\footnotesize{\footnotesize\caption{$T_c$ in units of $\sqrt{\rho_0}$ 
as a function of $N$ for $g(T=0) = 0.1$ (solid line).
For comparison results from 
\cite{ChristofNickfiniteT} (squares), and the result of an expansion in $1/N$
are also given.
Our results are very well reproduced by the naive perturbative result
(\ref{pertTc}) for all $N$.}}}
\end{minipage}
\end{center}
\end{figure}
\renewcommand{\baselinestretch}{1.2}\normalsize
For comparison, the dashed line shows the result that is obtained in leading
order in $1/N$ 
\cite{DolanJackiw}
\bea
T_c = \sqrt{\frac{24}{N}} \sqrt{\rho_0} +
{\mathcal{O}}\left(\frac{1}{N^2}\right)
\label{largeNTc}
\eea
Naive perturbation theory yields (\ref{pertTc})
which is consistent with our results on the level of $1-2 \%$ for the small
value of $g_{T=0} = 0.1$ choosen here (see figures 5
and 8 for a comparison of $T_c$ with (\ref{pertTc}) as a function of the coupling).
Also given are the results from the ERG in the imaginary-time formulation
\cite{ChristofNickfiniteT} as squares.
Again we observe good agreement as was already seen in figures 5 and 8.
It is interesting that we can also study the limit $N \rightarrow 0$
straightforwardly.

\section{Conclusions and outlook}

In the present paper we have discussed the finite-temperature
phase-transitions of self-interacting scalar theories with at
vanishing temperature spontaneously broken $O(N)$-symmetry.
We have used an implementation of the Wilsonian renormalization
group-approach for field theory in thermal equilibrium in the
Schwinger-Keldysh (CTP)-formulation
\cite{DAP1}.

We have reviewed and discussed this ``thermal renormalization group''
(TRG)-approach in section 2.
The approach has a number of advantages as compared to the more
common RG-approach in the framework of imaginary-time thermal field
theory:
It allows for the direct computation of real-time observables in field
theory at finite temperature, avoiding the need for analytical
continuation.
Thus quantities like thermal damping rates and the like are easily
accessible and first results may be found in the literature
\cite{Massimo}.
As the imaginary-time RG approach, the method is non-perturbative and allows for the treatment of
situations in which perturbation theory fails, which is typically the
case near second order phase-transitions or in strongly coupled
theories.

Since the real-time formulation of thermal field theory clearly
distinguishes thermal and quantum-fluctuations and the TRG treats only
the thermal fluctuations, one obtains a direct connection between the
physical couplings of the theory at vanishing and at non-vanishing
temperature and there is no need to take into account zero-temperature
renormalization scheme dependencies.
Furthermore the TRG manifestly respects gauge-invariance.

Finally, after a straightforward generalization of the formalism given
in section 2, the TRG may easily be applied to field theory at finite
density (chemical potential) or to theories with a non-thermal
density matrix.
The method thus offers a nonperturbative approach for the study of a
number of interesting questions in field theory in a hot and possibly
dense environment.

The main part of the present work is concerned with the application of
this formalism to $O(N)$-symmetric scalar theories in $3+1$ dimensions
at temperatures near to the critical one.
We have given extensive discussions of the cases $N=1$ and $N=4$,
being especially interesting for a number of reasons.
The case $N=1$ in three dimensions corresponds to the well known
Ising-model.
The electroweak sector of the standard model is in the
universality-class of the Ising model if the zero-temperature Higgs
mass is at its critical value $\sim 70$~GeV
\cite{EWPTsecondorder}.

We have given the critical temperature in terms of the scalar
self-coupling at $T=0$ and found good agreement with results found
from the exact renormalization group in the imaginary-time formulation
\cite{ChristofNickfiniteT}
(figure 5).
The results disagree with the values found in the
environmentally friendly RG
\cite{Chrisetal1}
but are consistent with results from the auxiliary mass method
\cite{Satoetal},
where the latter yields results that are basically identical to the
perturbative values for couplings up to $g = 1/3$  (figure 3 of the last
reference of 
\cite{Satoetal}).
Although the approach of 
\cite{Chrisetal1} differs from the one used in the present work,
the discrepancies of the results on the critical temperature are somewhat
surprising and should be further understood.

We have then at length discussed the critical behavior of the theory,
focusing on the universal aspects encoded in critical exponents,
amplitude ratios and the scaling equation of state (tables 1 and 2,
figures 6 and 7).
Whereas the exponents found in the present work are not as accurate as
the results found from Borel-resummed $\epsilon$-expansion and other
methods working directly in the three-dimensional effective model, the
scaling equation of state compares very favorably with results from
Monte-Carlo simulations in three dimensions.
This demonstrates the power of the TRG and also demonstrates
explicitly in a nonperturbative context how the critical behavior of
the finite temperature theory is governed by the three-dimensional
Euclidean model.

After the discussion of $N=1$ we have turned to the case $N=4$, which
is presumably in the same universality-class as the chiral
phase-transition in the case of 2 massless flavors.
We find a similar situation as in the case of $N=1$: The critical
temperatures are compatible with the results from the ERG in
imaginary-time
\cite{ChristofNickfiniteT} (figure 8).
For this case, results from the environmentally friendly RG or the
auxiliary mass method are not available.

Concerning the critical behavior, we again find a scaling form of the
equation of state in very good agreement with the results from the
exact renormalization group as given in 
\cite{JNC2} (figure 9)
but a somewhat larger difference in the critical exponents and
amplitude-ratios (tables 3 and 4).
We have argued that this situation should improve going to higher
order in the derivative-expansion.

Finally we have given the critical temperature as a function of $N$
for a large range of $N$, using a small value of the zero-temperature coupling
$g_{T=0} = 0.1$ (figure 10).
We are able to study the limit $N\rightarrow 0$, which is of interest for
statistical physics and find results consistent with the values obtained from
the exact renormalization group in the imaginary-time formulation and also with
naive perturbation theory\footnote{One should note that perturbation theory
beyond the leading term fails to reproduce the dependence of the critical
temperature on the coupling constant
\cite{Chrisetal1}.}.

Altogether the TRG has proven a useful and flexible tool in the study
of field theory at finite temperature.
It allows for the study of universal and non-universal quantities and
may be extended to theories involving gauge-fields
\cite{DAP2}
and fermions or to situations at finite density or more general
density-matrices.
It offers a nonperturbative way of studying dimensional reduction and
furthermore allows for the investigation of real-time quantities and
questions related to the dynamics of systems at finite temperature.

\bigskip

{\bf{Acknowledgments:}} We would like to thank Daniel Litim, Johannes
Manus, Massimo Pietroni, Chris Stephens and Michael Strickland for helpful
discussions. B.B. was supported by 
the "Sonderforschungsbereich 375-95 f\"ur
Astro-Teilchenphysik" der Deutschen Forschungsgemeinschaft, J.R. acknowledges
support by a Promotionsstipendium des Freistaates Bayern.

\end{document}